\documentclass[%
 reprint,
 superscriptaddress,
 amsmath,amssymb,
 aps,
pra,
 floatfix,
]{revtex4-2}

\usepackage{graphicx}% Include figure files
\usepackage{dcolumn}% Align table columns on decimal point
\usepackage{bm}% bold math
\usepackage{hyperref}% add hypertext capabilities
\usepackage[usenames,dvipsnames]{color}
\usepackage{colortbl}
\usepackage[english]{babel}
\newcommand{\DEL}[1]{}

\begin{document}

%\preprint{APS/123-QED}

\title{Robust strong-field theory model for ultrafast electron transport \\ through metal-insulator-metal tunneling nanojunctions}% Force line breaks with \\

\author{Boyang Ma}
\affiliation{Department of Physics, Technion---Israel Institute of Technology, Haifa 32000, Israel}
\affiliation{Solid State Institute, Technion---Israel Institute of Technology, Haifa 32000, Israel}
\affiliation{The Helen Diller Quantum Center, Technion---Israel Institute of Technology, Haifa 32000, Israel}

\author{Michael Kr\"uger}
\altaffiliation{Corresponding author: krueger@technion.ac.il}
\affiliation{Department of Physics, Technion---Israel Institute of Technology, Haifa 32000, Israel}
\affiliation{Solid State Institute, Technion---Israel Institute of Technology, Haifa 32000, Israel}
\affiliation{The Helen Diller Quantum Center, Technion---Israel Institute of Technology, Haifa 32000, Israel}

\date{\today}

\begin{abstract}
Ultrafast science studies the dynamics of electrons in matter with extreme temporal precision, typically in the attosecond and femtosecond time domain. Recent experimental and theoretical progress has put metal-insulator-metal (MIM) tunneling nanojunctions in the spotlight of ultrafast science. Waveform-controlled laser fields can induce ultrafast currents in these junctions, opening the door to petahertz electronic operation and attosecond-scale scanning tunneling microscopy (STM). Inspired by our strong-field model for attosecond tunneling microscopy [Boyang Ma and Michael Kr\"uger, Phys.~Rev.~Lett.~133, 236901 (2024)], here we generalize our model to MIM nanojunctions. We introduce several refinements and corrections, accounting for the image potential inside the gap and boundary effects. Moreover, we also find that the Keldysh parameter, which is a hallmark parameter for ultrafast light-matter interactions, alone is insufficient for describing the physics in thin MIM nanojunctions. We introduce a new parameter $\zeta$ that accounts for the effects of the limited size of the junction and provide several interpretations of the parameter that shed new light on the complex physics in the light-driven junction. Most strikingly, we find that for $\zeta > 1$ photon-assisted tunneling is dominating the ultrafast electron transport across the junction, regardless of the value of the Keldysh parameter. Here one or more photons are absorbed and the electron undergoes static tunneling through the barrier. $\zeta < 1$ is required for a regime in which the laser field and the Keldysh parameter dominate the transport. We also discuss the three-step model of electron transport across the nanojunction in the adiabatic regime, the energy cutoff of the transported electrons and the carrier-envelope phase control of the net current. Our theory model provides a rich toolbox for understanding and predicting the physics of ultrafast MIM nanojunctions.
\end{abstract}

\maketitle

%\tableofcontents

\section{\label{sec:level1}INTRODUCTION}

In recent years, attosecond science has seen tremendous development and has been applied in fields beyond its initial realm of atomic and optical physics~\cite{Krausz2009,Corkum2007}, such as molecular physics~\cite{Lepine2014,Palacios2020,Nisoli2017} and condensed-matter phyiscs~\cite{Cavalieri2007,Vampa2015,Ghimire2011}. High-harmonic generation (HHG) in gases and solid-state systems and photoelectron emission are the foundational effects in attosecond science~\cite{Corkum1993,Vampa2014,weissenbilder2022}. The unique properties of attosecond time scales make them essential for studying electron dynamics in atomic-scale matter. Such studies require ultrafast observation that does not disrupt the coherent evolution of the electrons in matter, alongside atomic resolution on $\mathring{\mathrm{A}}$ngstrom scales~\cite{Krausz2014}. Recently, the attosecond time scale has been introduced in electron microscopy approaches, such as transmission electron microscopy~\cite{Feist2015,Nabben2023}. The electron naturally possesses a short de Broglie wavelength, enabling a high spatial resolution in microscopy. By embedding attosecond time scales into electron pulses, it is possible to achieve high spatial-temporal resolution at the same time.

However, electron wavepackets stretch in time as they propagate due to the dispersion relation~\cite{Eldar2022}. To address this issue, a straightforward approach is to bring the detector (or sample) closer to the electron source. The most extreme case is a metal-insulator-metal (MIM) nanojunction, consisting of two metal contacts separated by a nanojunction filled with an insulating material, for example a vacuum. In history, Giaever and Megerle first designed this structure to verify the BCS theory of superconductivity~\cite{giaever1961}. The paradigm of MIM nanojunctions was used in the inception of the scanning tunneling microscopy (STM)~\cite{Binnig1982,Chen2021}. As an important MIM application, STM plays a crucial role in physics, chemistry, material science, and biology~\cite{Chen2021,Jung1997,Marti2012}. Due to the extremely thin junction, electron transfer between the two metal electrodes occurs via the tunneling effect. When a femtosecond laser field drives an MIM junction, it generates an ultrafast current that is well confined in both space and time~\cite{Rybka2016,Ludwig2020,Bionta2021,Garg2020}.  

Another paradigm in light-matter interactions involving MIM junctions is the strongly enhanced near-field caused by plasmonic effects in nanostructures which significantly amplify light–matter interactions~\cite{Schotz2019,Ciappina2017,Dombi2020}. This pronounced enhancement of the local field can exceed the amplitude of the incident laser by orders of magnitude. Moreover, the near-field is spatially confined at sharp nanometric protrusions and junctions, resulting in photoemission with a high degree of spatial coherence~\cite{Ehberger2015,Meier2018}. These two properties of the laser-assisted MIM nanojunction lead to several key consequences: they lower the threshold for the laser intensity, localize the electron burst, steer the electron motion, and create high-energy electrons~\cite{Dombi2020}. Furthermore, an intense few-cycle laser pulse can easily break the symmetry of the nanojunction while also introducing coherent control through the carrier-envelope phase (CEP)~\cite{Rybka2016}. In the most extreme case, single-cycle terahertz pulses have been shown to achieve extreme symmetry breaking. For instance, Cocker \textit{et al.} coupled an STM junction with a single-cycle terahertz beam, achieving manipulation of the tunneling current~\cite{Cocker2013} and enabling direct observations of molecular vibrations with picosecond resolution~\cite{Cocker2016}. Vedran \textit{et al.} and Katsumasa \textit{et al.} controlled the tunneling current in THz-STM by switching the polarization and CEP, respectively~\cite{Jelic2017,Yoshioka2016}. Garg and Kern were the first to observe attosecond currents in STM using a near-infrared laser~\cite{Garg2020}, with application to molecular charge state oscillations~\cite{Garg2022}. Besides ultrafast STM, a wealth of research on other electronic devices has emerged. Wenqi \textit{et al.} studied quantum effects on plasmonic resonances in two separated metallic particles~\cite{Zhu2016}. Savage \textit{et al.} observed tunneling between two tips using a white-light laser~\cite{Savage2012}. In a bow-tie antenna junction, Rybka \textit{et al.} used a single-cycle near-infrared pulse to achieve CEP control of the current~\cite{Rybka2016}. Ludwig \textit{et al.} demonstrated electron transport in a 6-nm gap bow-tie structure~\cite{Ludwig2020}. All nanodevices in these studies can be classified as laser-assisted MIM junctions.

Despite experimental achievements, a robust strong-field theory with an accurate description has remained elusive, particularly for ultrathin junctions with widths of approximately 1\,nm, such as STM junctions. Static electron tunneling cannot be neglected in these junctions, even in the absence of lasers. Furthermore, unlike THz-STM, a junction exposed to an intense few-cycle near-infrared laser is far from an adiabatic condition. The laser cycle duration is comparable to the intrinsic tunneling time of the current, meaning that electron dynamics must be considered~\cite{Sainadh2020}. In our previous work, we introduced a time-dependent theory based on the strong-field approximation (SFA), incorporating a novel three-step model for laser-assisted STM~\cite{Ma2024}. This theory provides a semi-classical picture for photoemissions in STM junctions, distinguishing them into the multiphoton regime and the laser-induced tunneling (adiabatic) regime. We indicate that photoemissions in the STM can also be characterized by the well-known Keldysh parameter. However, our model has serious drawbacks. For instance, the strong-field approximation neglects the image potential and a range of boundary effects. A cutoff energy of the electron spectrum that encompasses all emission regimes is still unknown, and the electron dynamics in each regime require further discussions.

In this paper, we establish a strong-field theory for ultrathin MIM nanojunctions, addressing the drawbacks of our previous theory approach and providing more accurate calculation results and predictions. The key advancements are outlined below.

(1) We derive the tunneling amplitude in detail, obtaining a time-dependent version of Bardeen's tunneling matrix element consistent with the tunneling effect. As a correction to the SFA, we introduce the Van Vleck propagator as a replacement for the Volkov propagator. Unlike the Feynman path-integral approach~\cite{Tannor2007}, the Van Vleck propagator is constructed using classical trajectories, allowing for the selection of specific paths and the incorporation of the image potential.

(2) The freed electron is constrained by the narrow width of the junction, preventing infinite acceleration by the laser as seen in high-harmonic generation (HHG) or larger nanojunctions. Electrons bouncing off the boundaries may not only transmit, but also scatter, causing energy accumulation and affecting the overall tunneling probability~\cite{Kruger2011}. In this work, we incorporate trajectory selection into the propagator for continuum states, filtering out unphysical trajectories and resolving the issue of unexpected ponderomotive energy accumulation.

(3) We propose a correction approach for the transmission process that reduces continuity and scattering errors, enhances the overall reliability of the calculations, and ensures more consistent and accurate results.

(4) We introduce a three-step model that incorporates the image potential to describe the photoemission mechanism in MIM junctions. Semi-classical trajectories, which account for tunneling effects, provide comprehensive description of the emission process.

(5) The cutoff energy, representing the maximum energy allowed by classical dynamics~\cite{Lewenstein1994,Ma2024}, has been extensively studied in the adiabatic regime. However, it remains less explored in the multiphoton and non-adiabatic regimes. In this work, we propose a unified definition of cutoff energy that is applicable across all regimes, along with a physical interpretation specific to each regime.

(6) We introduce a new parameter, $\zeta$, which governs the conditions under which strong-field photoemission occurs in MIM systems. Our findings demonstrate that the Keldysh parameter $\gamma$ alone is insufficient to fully describe the interactions in narrow nanojunctions. Instead, a characterization of the effective photoemission mechanism requires considering both $\gamma$ and $\zeta$. The physical interpretations and implications of this new parameter are explored in detail.

(7) We investigate the conditions required for CEP modulation in laser-assisted nanojunctions. This widely used technique in attosecond physics enables a coherent breaking of the spatial symmetry and precise control of attosecond electron trajectories. Through simulations, we explore and identify optimized conditions for effective CEP modulation.

This paper is organized as follows. In Sec.~\ref{Sec2}, we describe the basic approach to derive the tunneling amplitude without relying on approximations. In Sec.~\ref{Sec3}, the SFA is applied to the tunneling amplitude, incorporating the effects of the image potential and boundary constraints. The van Vleck propagator is introduced into our theory and boundary corrections are developed. In Sec.~\ref{Sec4}, we apply the saddle-point technique. Three-step model, semi-classical trajectories, classifications of photoemission mechanisms, the cutoff energy, and the new parameter $\zeta$ are obtained. In Sec.~\ref{Sec5}, we conclude with a summary of our findings and provide an outlook on future research directions.

\section{\label{Sec2}BASIC APPROACH}

\subsection{\label{sec2.1}Time-dependent metal-insulator-metal junction model}

\begin{figure*}[t!]
\centering
\includegraphics[clip, trim=1cm 0cm 2cm 0cm, width=1\linewidth]
{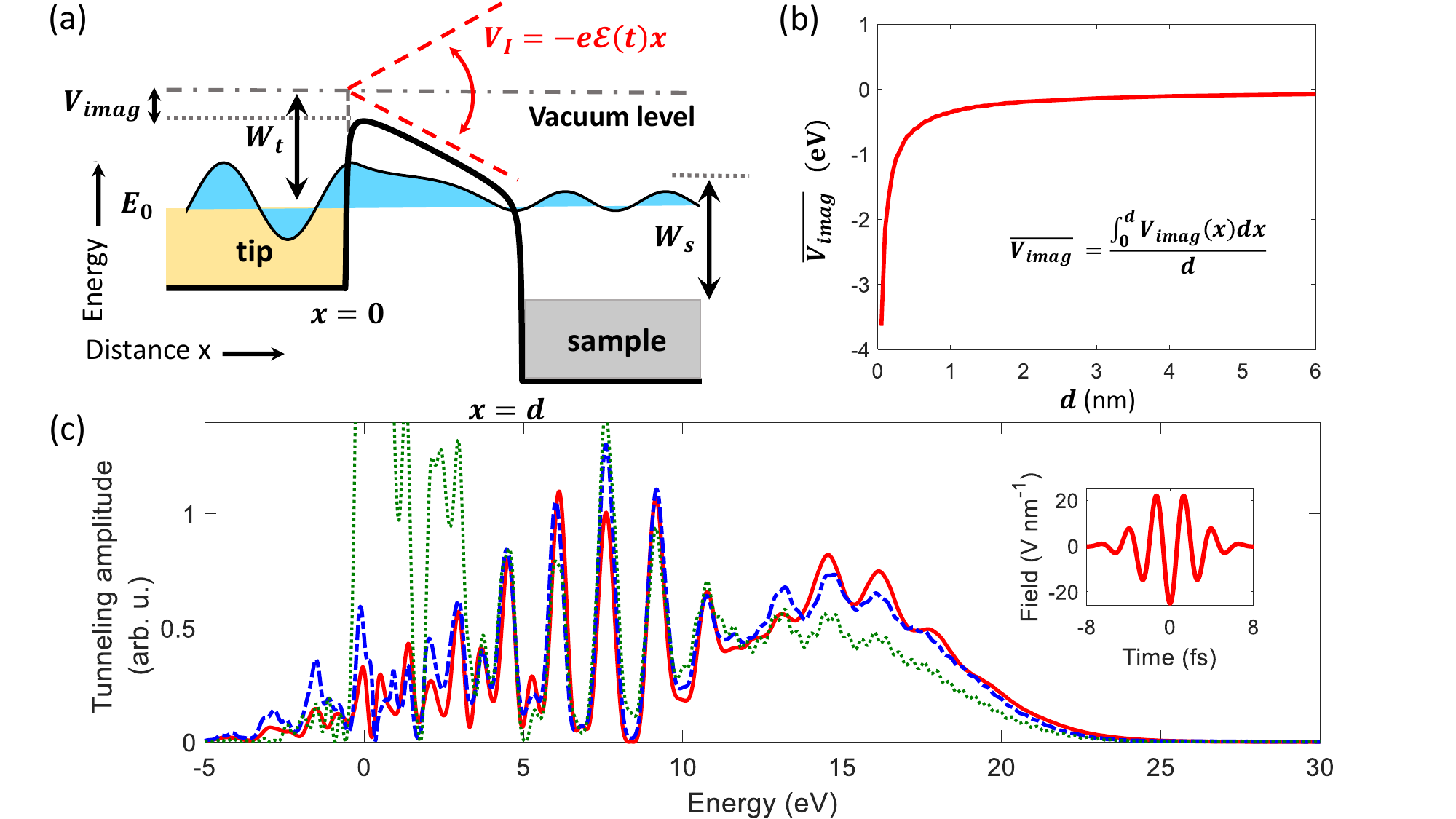}
\caption{Strong-field theory for a MIM nanojunction. (a) A schematic sketch of a laser-driven MIM nanojunction. $W_\mathrm{t}$: tip work function, $W_\mathrm{s}$: sample work function, ${\cal E}(t)$: laser electric field. (b) The variation of the spatially averaged image potential $\overline{V_{\mathrm{imag}}}$ with the junction width $d$. (c) The tunneling electron spectrum calculated by the TDSE (red solid line), SFA theory (green dotted line), and SFA theory with corrections (blue dashed line).  For this calculation, the junction width is $d = 1$\,nm, the laser pulse with central wavelength 830\,nm has a duration of 6\,fs (full width at half maximum, see inset for the field waveform) and the field strength $F$ is $25\,\mathrm{V\,nm}^{-1}$. The material for tip and sample is gold ($W_\mathrm{t} = W_\mathrm{s} = 5$\,eV), the tip electron is initially at the Fermi energy $E_0 = -W_\mathrm{t} = -5$\,eV, and no bias potential $U_{\mathrm{s}}$ is applied.}
\label{Fig1}
\end{figure*}

An ultrathin MIM nanojunction is formed by two conductive metals separated by an insulating gap, as shown in Fig.~\ref{Fig1}(a). Following the convention used in STM, we label the left contact as the tip and the right contact as the sample. We assume that the insulator gap is vacuum. This vacuum junction is so small that the electronic wave functions within the tip and sample can tunnel through it with non-negligible probability, leading to static tunneling currents~\cite{Simmons1963,Scholl2013}. By applying a bias voltage to the system (with both metals possessing Fermi energies at the same level), the junction symmetry is broken, causing the static tunneling current to flow in only one direction.

%
% FIGURE 1
%

We assume that electrons move freely in the metals and possess an effective mass identical to the rest mass in vacuum~\cite{Yalunin2011}. The Fermi energies in the tip and sample are defined as $E_\mathrm{F, t}$ and $E_\mathrm{F, s}$, respectively. Their work functions are defined as $W_\mathrm{t}$ and $W_\mathrm{s}$, respectively~\cite{Yoshioka2016, Jelic2017}. The field-free junction potential in the length gauge is given by
\begin{equation}\label{eq2.1.1} V_{0}(x) = \begin{cases} -(E_\mathrm{F, t}+W_\mathrm{t}), &  x < 0 ,\\ V_\mathrm{imag}(x)-e(\varphi+U_\mathrm{s}) x/d, & 0 \leqslant x \leqslant d,\\ -(E_\mathrm{F, s}+W_\mathrm{s}+e\varphi+e U_\mathrm{s}), & x > d, \end{cases}
\end{equation}
where $e = -|e|$ is the charge of the electron and $V_\mathrm{imag}(x)$ is the spatially dependent image potential. $\varphi=(W_\mathrm{t}-W_\mathrm{s})/e$ is the contact potential difference, which is also known as Volta potential, and $U_\mathrm{s}$ is the static bias voltage applied to the junction. Here, the tip boundary is set at $x=0$ and the sample boundary is set at $x=d$. According to the Simmons theory, the image potential inside the gap is described as~\cite{Yoshioka2016, Simmons1963}
\begin{equation}\label{eq2.1.2}
V_\mathrm{imag}(x)=\left(-\frac{e^2}{8\pi\varepsilon}\right)\bigg[\frac{1}{2x}+\sum^{\infty}_{n=1}{\left\{\frac{nd}{(nd)^2-x^2}-\frac{1}{nd}\right\}}\bigg].
\end{equation}
Here, the permittivity $\varepsilon$ of the vacuum gap is $1$. Figure~\ref{Fig1}(b) shows how the averaged image potential $\overline{V_{\mathrm{imag}}}$ varies with the junction width $d$. It approaches zero when the width exceeds 1\,nm.

When irradiated with a laser field, the tip and sample are considered ideal metals, ensuring that the laser field is perfectly screened. We assume that the polarization of the laser electric field is along the $x$-axis, so the interaction in the length gauge is given by
\begin{equation}\label{eq2.1.3} 
V_\mathrm{I}(x,t) = \begin{cases} 0, &  x < 0 ,\\ -e{\cal E}(t)x, & 0 \leqslant x \leqslant d,\\ -e{\cal E}(t)d, & x > d \end{cases}. \end{equation}
where ${\cal E}(t)$ is the electric field of the laser. The dipole approximation is justified here because the width of the vacuum junction is much smaller than the laser wavelength~\cite{Garg2021}. The full potential $V(x,t)=V_{0}(x)+V_{\mathrm{I}}(x,t)$ combines the static potentials with the electric dipole and image potential and  describes photoemission, charge transport, and barrier reduction effects, such as the Schottky effect~\cite{ Schottky1914, Schenk2010}. The electronic wavefunctions in the laser-assisted nanojunction are governed by the time-dependent Schr\"odinger equation (TDSE)
\begin{equation}\label{eq2.1.4}
i\hbar\frac{\partial}{\partial t}\Psi(x,t)=-\frac{\hbar^2}{2m}\frac{\partial^2}{\partial x^2}\Psi(x,t)+V(x,t)\Psi(x,t),
\end{equation}
where $\hbar$ is the reduced Planck constant and $m$ is the mass of the electron. The tunneling wavefunctions in the presence of the laser field can be obtained by solving this TDSE.

\subsection{\label{sec2.2}Current conservation}
In practice, experimental observations in MIM structures are performed by detecting currents. The wavefunctions traversing the vacuum gap must satisfy the continuity condition
\begin{equation}\label{eq2.2.1}
\frac{\partial}{\partial t}\vert\Psi\vert^2+\nabla\cdot J=0.
\end{equation}
Here $\Psi$ is the electronic wavefunctions at arbitrary time $t$. $J$ is the corresponding current density. The rapid changes induced by the laser field are too fast to be recorded due to the slow electronic response. Hence, the laser-induced tunneling (LIT) current recorded in experiments is averaged over the comparatively long time scale of conventional electronics~\cite{Garg2020}.

We define that the interaction starts at $-\infty$ and ends at $+\infty$. The entire system relaxes and returns to its original state. It is easy to obtain the flux conservation of currents through Eq.(~\ref{eq2.2.1}) under the relaxation condition, that $\int_{-\infty}^{+\infty}J_{s}\;dt\,=\int_{-\infty}^{+\infty}J_{t}\;dt\,$. Here, $J_{t}$ and $J_{s}$ are used to represent the current from the tip and the sample, respectively.

With the laser interaction, the wavefunctions in the tip and sample are perturbed and redistributed. Wavefunctions $\vert\Psi_{t}\rangle$ and $\vert\Psi_{s}\rangle$ which are initially located in the tip and the sample are driven out by the laser, bifurcating into the tunneling wavefunctions $\vert\Psi_{tT}\rangle$ and $\vert\Psi_{sT}\rangle$ in the sample and the tip, respectively. Based on Eq.~\ref{eq2.2.1} and the conservation constraint, we obtain the accumulated current ${\cal J}$ at the final time as follows (for detailed derivations see Appendix~\ref{A}): 
\begin{equation}\label{eq2.2.2}
{\cal J}=\langle \Psi_{tT} \vert\Psi_{tT}\rangle-\langle \Psi_{sT} \vert\Psi_{sT}\rangle.
\end{equation}

Eq.~\ref{eq2.2.2} shows that the net current is only related to the tunneling probabilities from the two sides. Assume the field-free eigenfunction inside the sample (tip) at energy $E$ is $\vert\psi_{sE}\rangle$ ($\vert\psi_{tE}\rangle$), the current $\cal J$ can be rewritten as
\begin{eqnarray}\label{eq2.2.3}
{\cal J}&=&\sum_{E}[\langle \Psi_{tT} \vert\psi_{sE}\rangle\langle\psi_{sE}\vert\Psi_{tT}\rangle-\langle \Psi_{sT} \vert\psi_{tE}\rangle\langle\psi_{tE}\vert\Psi_{sT}\rangle]\nonumber\\
&=&\sum_{E}(\vert M_{tE}\vert^2-\vert M_{sE}\vert^2),
\end{eqnarray}
where the completeness of eigenfunctions has been used. $M_{tE}$ and $M_{sE}$ are tunneling amplitudes for $\vert\Psi_{t}\rangle$ and $\vert\Psi_{s}\rangle$, respectively.

\subsection{\label{sec2.3} Tunneling amplitude}

In the preceding section and Appendix~\ref{A}, we relate the net current to the tunneling amplitude and demonstrate the independence of the currents between the two contacts. Due to the geometric structure and the polarization of the laser, it is straightforward to demonstrate that $\vert\Psi_{sT}\rangle$ can be also solved in the same way as $\vert\Psi_{tT}\rangle$, only switching the conditions from the tip to the sample (see Appendix ~\ref{B} and Refs.~\cite{Luo2021,Ma2024}). Hence, in the following, we focus only on the evolution of $\vert\Psi_{t}\rangle$ and drop the subscripts. 

Before the laser interaction, the MIM is in a static condition and its initial wavefunction in the energetic state $E_0$ is $\vert\Psi_{0}\rangle$ (from the tip to the sample) as shown in Fig.~\ref{Fig1}(a). When the laser is switched on, the wavefunction evolves with the time and its tunneling amplitude at the final time is
\begin{equation}\label{eq2.3.1} M_{E}=\langle\psi_{E} \vert\chi_{s} U(\infty,-\infty) \vert\Psi_{0} \rangle,
\end{equation}
where $U(t_2,t_1)$ denotes the evolution operator with $t_2>t_1$. $\vert\psi_E\rangle$ is an eigenfunction of the sample in the energetic state $E$. We use the operator $\chi_{\mathrm{s}}$ to select spatial integration only in the sample domain. To elucidate the processes of ionization and transmission, we use the Dyson equation to expand the evolution operator to the second order~\cite{Milosevic2006, Ivanov2005}.
\begin{eqnarray}\label{eq2.3.2}
&&U(\infty,-\infty)=U_{0}(\infty,-\infty)\nonumber\\
&&+\left(-\frac{i}{\hbar}\right)\int_{-\infty}^{\infty}U(\infty,t_{1})\chi_{\mathrm{s}}V_\mathrm{I}(t_{1})U_{0}(t_{1},-\infty)\;dt_{1}\,\nonumber\\
&&+\left(-\frac{i}{\hbar}\right)\int_{-\infty}^{\infty}U_\mathrm{Is}(\infty,t_{1})\chi_{\mathrm{gap}}V_\mathrm{I}(t_{1})U_{0}(t_{1},-\infty)\;dt_{1}\,\nonumber\\
&&+\left(-\frac{i}{\hbar}\right)^{2}\int_{-\infty}^{\infty}\int_{-\infty}^{t_{2}}U(\infty,t_{2})[\chi_{\mathrm{t}}+\chi_{\mathrm{s}}][V(t_{2})-V_\mathrm{Is}(t_2)]\nonumber\\
&& \ \ \ \times \, U_\mathrm{Is}(t_{2},t_{1})\chi_{\mathrm{gap}}V_\mathrm{I}(t_{1})U_{0}(t_{1},-\infty)\;dt_{1}dt_{2}\,.
\end{eqnarray}
where $U_{0}(\cdot,\cdot)$ is the evolution operator for the $V_{0}$, $\chi_{\mathrm{gap}}$ and $\chi_{\mathrm{tip}}$ are selection operators applied in the gap and tip, respectively, and $U_{\mathrm{Is}}(\cdot,\cdot)$ is the evolution operator for $V_\mathrm{Is}$, which is an interaction with partial static potentials:
\begin{equation}\label{eq2.3.3} 
V_\mathrm{Is}(t) = \begin{cases} 0, &  x < 0 ,\\ V_\mathrm{imag}-ex[{\cal E}(t)+(\varphi+U_\mathrm{s})/d], & 0 \leqslant x \leqslant d,\\ -e[{\cal E}(t)d+\varphi+U_\mathrm{s}], & x > d. \end{cases} 
\end{equation}
The subscript ``$\mathrm{Is}$'' represents the time-dependent interaction with the static potentials. The first term of Eq.~\ref{eq2.3.2} describes the unperturbed evolution, while the second term represents the phase modulation after the static tunneling. Both terms are unrelated to the laser-induced charge transport and can be eliminated by subtraction in Eq.~\ref{eq2.2.3} and by applying periodic boundary conditions to the sample. Substituting the rest into Eq.~\ref{eq2.3.1}, we obtain the tunneling amplitude $M_{E}$ as
\begin{eqnarray}\label{eq2.3.4}
M_{E}&=&\langle\psi_{E}\vert\chi_\mathrm{s}\vert\Psi_\mathrm{Is}\rangle+\left(-\frac{i}{\hbar}\right)\int_{-\infty}^{\infty}\langle\psi(t_1)\vert[\chi_\mathrm{t}+\chi_\mathrm{s}]\nonumber\\
&&\times[V(t_{1})-V_\mathrm{Is}(t_{1})]\vert\Psi_\mathrm{Is}(t_{1})\rangle\;dt_{1}\, ,
\end{eqnarray}
where $\langle\psi(t_{1})\vert=\langle\psi_{E}\vert\chi_{\mathrm{s}}U(\infty,t_{1})$ and $\vert\Psi_\mathrm{Is}(t)\rangle=(-\frac{i}{\hbar})\int_{-\infty}^{t} U_\mathrm{Is}(t,t_{1})\chi_{\mathrm{gap}}V_\mathrm{I}(t_{1})\vert\Psi_{0}(t_{1})\rangle\;dt_{1}\,$. They are wavefunctions that evolve inside the junction and the sample, respectively. The potentials in Eq.~\ref{eq2.3.4} can be replaced with the Schr\"odinger equation (for detailed derivations, see Appendix~\ref{C}) and are further simplified as

\begin{equation}
\begin{split}
M_{E}= \frac{i\hbar}{2m}\int_{-\infty}^{\infty} \bigg[\Psi_{\mathrm{Is}}(x,t)\frac{\partial}{\partial x}\psi^{*}(x,t)\\
-\,\psi^{*}(x,t)\frac{\partial}{\partial x}\Psi_{\mathrm{Is}}(x,t)\bigg]\bigg\vert^{x=d}_{x=0}\;dt,
\label{eq2.3.5}
\end{split}
\end{equation}
where $\Psi_\mathrm{Is}(x,t)$ and $\psi^{*}(x,t)$ are $\langle x\vert\Psi_\mathrm{Is}(t)\rangle$ and $\langle\psi(t)\vert x\rangle$, respectively. The notation $[...]\big\vert^{x=d}_{x=0}$ at the end of Eq.~\ref{eq2.3.5} stands for the subtraction of the term inside the brackets evaluated at $x=0$ from the term evaluated at $x=d$.

$\Psi_{\mathrm{Is}}(x,t)$ is the time-propagated wave function driven by the laser, which is generated from $\left| \Psi_0 \right\rangle$ inside the gap.  $\psi^{*}(x,t)$ is the complex conjugate of the eigenfunction in the sample region propagated to a specific time. Eq.~\ref{eq2.3.5} reveals the connection between these wavefunctions located on either side of the sample boundary. It has a typical form of Bardeen's tunneling matrix element~\cite{Bardeen1961}, which is a standard theoretical model for the static STM~\cite{Chen2021}. In our $M_E$, the matrix element becomes time-dependent. Equation~\ref{eq2.3.5} also includes the tip-vacuum boundary at $x=0$. The field-driven $\Psi_{\mathrm{Is}}(x,t)$ in the gap can scatter at the tip boundary at $x=0$, leading to a rescattering effect~\cite{Kruger2011}. The tunneling amplitude inside the sample also depends on the flux through the tip boundary, which is a direct result of the current continuity. Equation~\ref{eq2.3.5} is a formal solution of the TDSE applied to MIM nanojunctions. Its evaluation requires a series of approximation methods, which we will discuss in the following.

%
% FIGURE 2
%
\begin{figure*}[t!]
\centering
\includegraphics[clip, trim=0cm 0cm 2cm 0cm, width=1\linewidth]{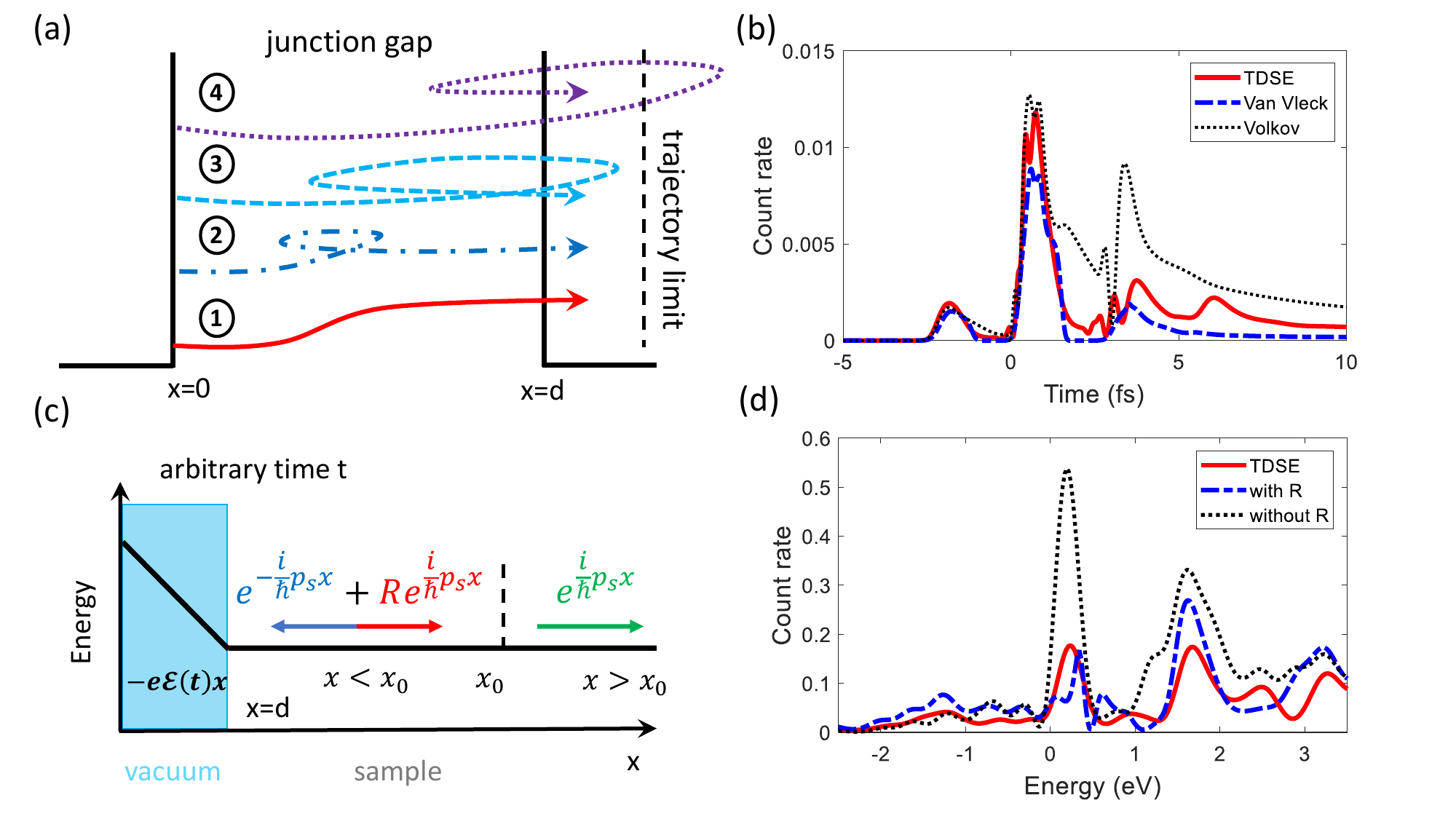}
\caption{Corrections to the SFA and comparison with exact TDSE solutions. (a) Possible electron trajectories in the junction. Trajectories (1) and (2) represent direct electrons and satisfy the physical boundary constraints. Trajectory (3) can be included in a rescattering approximation. Trajectory (4), however, is unphysical and must be excluded as it penetrates deeply into the (force-free) sample. (b) Comparison of the amplitude of $\Psi_{\mathrm{Is}}$ at the sample boundary ($x=d$) according to three methods, TDSE (solid red), Van Vleck (dashed blue) and Volkov (dotted black). The Volkov propagator without trajectory selection strongly deviates from the TDSE simulation after half a cycle (1.4 fs). (c) Illustration of the Green's function in the sample region. A wave generated at $x_{0}$ produces both a forward-propagating plane wave and a backward wave. A reflection at the sample boundary with an average amplitude $R$ is taken into account. (d) Comparison of the tunneling spectrum according to 3 approaches, TDSE (solid red), Van Vleck with reflection (dashed blue) and without reflection (dotted black). The curve without reflection correction exhibits an unphysical enhancement peak near zero energy. The simulation parameters are identical to those used in Fig.~\ref{Fig1}. }
\label{Fig2}
\end{figure*}

\section{\label{Sec3}STRONG-FIELD APPROXIMATION AND VAN VLECK PROPAGATION}
\subsection{\label{sec3.1}Strong-field approximation}

In the previous section, we derived the formula for the tunneling amplitude and established its connection to the currents. Eq.~\ref{eq2.3.5} requires finding the values of $\Psi_{\mathrm{Is}}(x,t)$ and $\psi^{*}(x,t)$ at the junction boundaries. Exact analytical expressions are unattainable for the complex interactions with the laser field. Therefore, we focus only on the dominant interactions in each process, neglecting weak perturbations -- a treatment known as the strong-field approximation~\cite{Ivanov2005, Kruger2012a,Ciappina2017,Dombi2020, Amini2019}. The SFA provides both qualitative and quantitative insight by emphasizing the dominant interactions. In the laser-assisted MIM junction, three main properties need to be considered in the context of the SFA:

(a) The image charge potential compared with the laser field is weak.

(b) The junction width is small compared with the electron's quiver length.

(c) The gradient of the interaction $-\partial V_{\mathrm{I}}(x,t)/\partial x$ is zero everywhere except within the junction.

However, unlike in the SFA for HHG~\cite{Ivanov2005, Lewenstein1994,Corkum2007} and ATI~\cite{Milosevic2006,Milosevic2002,Becker2018}, the field-free potential in the nanojunction cannot be neglected~\cite{Yoshioka2016}. A compromise approach is to remove the static field in the vacuum junction and independently evolve the wavefunctions within each region~\cite{Luo2021,Ma2024}. If the image potential and boundary effects are neglected, the electron driven by the laser field in the vacuum can be described with Volkov wavefunctions~\cite{Volkov1935}. The propagator corresponding to the Volkov wavefunctions is (Appendix~\ref{D}) 
\begin{equation}\label{eq3.1.1}
\begin{split}
\langle x_{2}\vert U_{V}(t_{2},t_{1})\vert x_{1} \rangle=\sqrt{\frac{m}{2\pi i \hbar(t_{2}-t_{1})}}e^{\frac{i}{\hbar}( p-eA(t))x_{2}}\\
\times e^{-\frac{i}{\hbar}( p-eA(t))x_{1}-\frac{i}{\hbar}\int_{t_{1}}^{t_{2}}\frac{(p-eA(\tau))^2}{2m}\;d\tau\,}.
\end{split}
\end{equation}
where $ p=\frac{\int_{t_{1}}^{t_{2}}{eA(\tau)}\;d\tau\,+m(x_{2}-x_{1}))}{t_{2}-t_{1}}$ is the canonical momentum and $A(t)=- \int^{t}_{-\infty} {\cal E}(\tau)\,d\tau$ is the vector potential. Classically interpreted, the Volkov propagator describes the wavefunction of a free electron traveling from $x_{1}$ to $x_{2}$ in a homogeneous laser field in the time frame between $t_{1}$ and $t_{2}$.

The approximation using the Volkov wavefunctions is justified for large junctions and has played a crucial role in many strong-field theories and attosecond experiments~\cite{ Lewenstein1994,Corkum2007,Milosevic2006,Kruger2011,Shafir2012}. However, due to the unconstrained wave propagation within the SFA, applying the Volkov function in the MIM junction can lead to unexpected energy and phase changes, strongly amplifying the ponderomotive effect where the electron performs a quiver motion~\cite{ Ma2024}.

Property (c), which is valid in the tip and sample domains, describes a force-free system where the classical charged particle can only move forward freely. Based on this property, the transmitted wavefunction in the sample corresponds to a traveling plane wave for a free electron, which is given by
\begin{eqnarray}
\label{eq3.1.2}
\psi^{*}(x,t)&=&\langle\psi_{E}\vert\chi_{s}U(\infty,t)\vert x \rangle\nonumber\\
&\approx&\sum_{E'}
\langle\psi_{E}\vert\chi_{s}\vert\psi_{E'}\rangle\langle\psi_{E'}(t)\vert x \rangle\nonumber\\
&=&\langle\psi_{E}(t)\vert x \rangle\nonumber\\
&=&e^{-\frac{i}{\hbar}(p_{s} x-E t-eA(t)d)}.
\end{eqnarray}
where $p_{s}$ is the momentum of the transmitted electron for the eigenstate $E$. In the sample, we have $p_{s}=\sqrt{2m[E-V_{0}(x)]},\ x>d$. Equation~\ref{eq3.1.2} drops the backward propagation solution so that it is valid only in the region far away from the boundary~\cite{Yalunin2011}.

Applying the two solutions above makes the tunneling amplitude in Eq.~\ref{eq2.3.5} analytically solvable. However, as discussed, they also introduce unexpected errors by neglecting the effects of the boundary. The Volkov wavefunction, without any trajectory constraints, is allowed to extend beyond the junction region. However, laser-driven motion outside the junction is unphysical and needs to be avoided in the theory. Moreover, the evolution inside the junction excludes any variation of the vacuum potential, such as the reduction of the vacuum potential by the image potential. As a result, the tunneling amplitudes may show unphysical resonant peaks in some cases. Corrections to the SFA are required.

In Fig.~\ref{Fig1}(c), we present a comparison between the SFA spectrum (green dotted line) and the TDSE spectrum (red solid line) to illustrate the shortcomings of the SFA. The TDSE is numerically solved using the Crank-Nicolson scheme~\cite{Yalunin2011,Kruger2011}. The use of unconstrained Volkov wavefunctions in the SFA introduces strong noise around 15\,eV, while the boundary artifacts in the transmission cause a strong divergence near zero energy.

\subsection{\label{sec3.2}Van Vleck propagator}

The Coulomb potential in laser-atom interactions poses a similar problem compared to the image potential in an MIM nanojunction. To incorporate the Coulomb potential into the strong-field description of HHG, the Coulomb-Volkov approximation and Eikonal approximation have been utilized~\cite{Reiss1994,Smirnova2008}. These approaches treat the long-range Coulomb potential as a weak perturbation, introducing corrections to both the phase and amplitude. In our case, we have to introduce a correction due to the image potential confined within the nanojunction. Therefore, a boundary constraint is necessary for the nanojunction. In this section, we introduce a new approach using the Van Vleck propagator, which accounts for classical trajectories in its formulation~\cite{Kay2013,Berry1989,Almeida1998}. The formal expression for the Van Vleck propagator is
\begin{eqnarray}
\label{eq3.2.1}
\begin{split}
\langle x_{2}\vert U_{\mathrm{Van}}(t_{2},t_{1})\vert x_{1} \rangle=&\sum_{cl}\frac{1}{\sqrt{2\pi i\hbar}}\bigg\vert \frac{\partial^2 S}{\partial x_{2}\partial x_{1}}\bigg\vert^{\frac{1}{2}}\\
&\times e^{\frac{i}{\hbar}S(x_{2},t_{2},x_{1},t_{1})-i\frac{\pi\nu}{2}},
\end{split}
\end{eqnarray}
where $\nu$ is the Morse index~\cite{Kay2013}, and action $S(x_{2},t_{2},x_{1},t_{1})$ is the time integral of the Lagrangian defined for each of the paths. The sum in the Van Vleck propagator is only over the classical paths (denoted as cl here), not over all possible virtual paths as in the Feynman path expression~\cite{Tannor2007}. For a movement from position $x_{1}$ to $x_{2}$ with a time-dependent momentum $p(t)$, the Lagrangian is written as
\begin{eqnarray}\label{eq3.2.2}
{\cal L}(x,t)=p(t)\frac{dx(t)}{dt}-\left\{\frac{p^{2}[x(t),t]}{2m}+V[x(t),t]\right\}.
\end{eqnarray}
Due to the interaction described by Eq.~\ref{eq2.1.3}, the laser field drives the electron only within the junction along classical trajectories (after the excitation from the initial state). Consequently, the initial position must lie within the junction ($0 \leq x_{1} \leq d$), and the trajectory cannot cross the boundaries before the electron reaches its final position at $x_{2} = d$, as illustrated by trajectories (1) and (2) shown in Fig.~\ref{Fig2}(a). The trajectories in the thin junction do not involve singular points; therefore, the Morse index is $\nu=0$. For an electron moving without any scattering, it must have a canonical momentum 
\begin{equation}\label{eq3.2.3.1}
 p=\frac{\int^{t_{2}}_{t_{1}}eA(\tau)\;d\tau\,+m(x_{2}-x_{1})}{(t_{2}-t_{1})},
\end{equation}
and a trajectory 
\begin{equation}\label{eq3.2.3}
x(t)=x_{1}+\int^{t}_{t_{1}}\frac{ p-eA(\tau)}{m}\;d\tau\,.
\end{equation}
Here, we neglect the force exerted by the image potential because it is weak compared to the time-dependent potential due to the strong laser field. Substituting all these into the Lagrangian, the action $S$ is finally expressed as
\begin{eqnarray}\label{eq3.2.4}
S&=&\int^{t_{2}}_{t_{1}}{\cal L}(x,t)\;dt\,\nonumber\\
&\approx&[p-eA(t_{2})]x_{2}-[p-eA(t_{1})]x_{1}\nonumber\\
&&-\int^{t_{2}}_{t_{1}}\left[\frac{( p-eA(t))^2}{2m}+V_{\mathrm{imag}}(x(t))\right]\;dt\,.
\end{eqnarray}
It should be noted that the above action only includes the ``direct electrons'' which never interact with boundaries ((1) and (2) in Fig.~\ref{Fig2}(a)), so the trajectory $x(t)$ has a constraint $0\leq x(t)\leq d$.

The derivative of the action $S$ is 
\begin{eqnarray}\label{eq3.2.5}
\frac{\partial^2 S}{\partial x_{1} \partial x_{2}}&=&\frac{\partial  p}{\partial x_{1}}+\int^{t_{2}}_{t_{1}}\frac{t-t_{1}}{t_{2}-t_{1}}\frac{\partial^2 V_{\mathrm{imag}}[x(t)]}{\partial x(t)^2}\frac{\partial x(t)}{\partial x_{1}}\;dt\,+...\nonumber\\
&\approx&-\frac{m}{t_{2}-t_{1}}.
\end{eqnarray}
Substituting it into the prefactor of Eq.~\ref{eq3.2.1}, the Van Vleck propagator for the nanojunction is finally expressed as
\begin{equation}
\label{eq3.2.6}
\begin{split}
\langle x_{2}\vert U_{\mathrm{Van}}(t_{2},t_{1})\vert x_{1} \rangle=\sum_{cl}\sqrt{\frac{m}{2\pi i \hbar(t_{2}-t_{1})}}e^{\frac{i}{\hbar}( p-eA(t_{2}))x_{2}}\\
\times e^{-\frac{i}{\hbar}( p-eA(t_{1}))x_{1}-\frac{i}{\hbar}\int_{t_{1}}^{t_{2}}[\frac{( p-eA(\tau))^2}{2m}+V_{\mathrm{imag}}(x(\tau))]\;d\tau\,}.
\end{split}
\end{equation}
If the trajectory constraint and the image potential are neglected, the result aligns with the Volkov propagator described in Eq.~\ref{eq3.1.1}. Therefore, the states described by the Van Vleck propagator can be referred to as spatially constrained Volkov states. The contribution of the image potential to the phase accumulates along the trajectory, in agreement with the conclusions of the Eikonal approximation~\cite{Smirnova2008,Smirnova2006}.

Up to this point, we have only considered direct electrons. In small junctions, additional paths arise from contributions by reflected electrons. This phenomenon, known as rescattering, has been observed at nanotips~\cite{Kruger2011} and in above-threshold ionization~\cite{Milosevic2006}. Electrons emitted from the tip are driven by the laser field, return to the parent metal surface after approximately $3/4$ of an optical cycle, and undergo elastic scattering~\cite{Kruger2011,Wachter2012}. However, in strong fields on the order of $10\,\mathrm{V/nm}$ or higher, most electrons traverse the small junction in less than half an optical cycle. As a result, the electrons do not quiver and rescattering is weak~\cite{Ludwig2020}. To incorporate rescattering within the junction, we employ a weak approximation. We consider trajectories that `gently' exceed the boundaries during their motion and quickly return ((3) in Fig.~\ref{Fig2}(a)). These trajectories can be interpreted as electrons that undergo reflection. It is important to emphasize that this is not a higher-order perturbation; instead, we use unconstrained trajectories to approximate the scattering effect. By relaxing the constraints, we obtain the following condition:
\begin{equation}
\label{eq3.2.7}
\begin{split}
x_{1}\geq& \frac{\int^{t_{2}}_{t_{1}}eA(\tau)\;d\tau\,+md}{m}\\
&-\bigg\lfloor \frac{\int^{t}_{t_{1}}eA(\tau)\;d\tau\,+md}{(t-t_{1})}\bigg\rfloor\frac{(t_{2}-t_{1})}{m}.
\end{split}
\end{equation}
where the notation $\lfloor...\rfloor$ denotes the minimum value during $t\in[t_{1},\ t_{2}]$. The trajectories of (1), (2), and (3) in Fig.~\ref{Fig2}(a) are included. Figure~\ref{Fig2}(b) illustrates the amplitude of $\Psi_{\mathrm{Is}}$ at the vacuum-sample boundary $x=d$. Compared to the TDSE, values obtained using the Volkov propagator deviate after half a cycle (1.4 fs), whereas wavefunctions derived with the Van Vleck propagator closely match the TDSE results.

\subsection{\label{sec3.3}Boundary effect }

The solution in Eq.~\ref{eq3.1.2} also requires a boundary correction. To incorporate the boundary effect into the approximated $\psi^{*}(x,t)$, we adopt the approach introduced by Yalunin \textit{et al.}~\cite{Yalunin2011}. Unlike the Fourier method, which matches the values at the boundaries~\cite{Faisal2006,Faisal2005}, this approach introduces a time-averaged reflection into the propagator. The propagator in the sample $\sum_{E'}\langle x_{2} \vert\psi_{E'}(t_{2})\rangle\langle\psi_{E'}(t_{1})\vert x_{1} \rangle$ can be interpreted as a matrix of forward plane waves and backward plane waves (see Fig.~\ref{Fig2}(c) for an illustration). There should be reflections for the backward waves after interacting with the sample boundary at $x=d$. Therefore, we introduce the boundary correction into $\psi^{*}(x,t)$:
\begin{eqnarray}
\label{eq3.3.1}
\psi^{*}(x,t)&=&\langle\psi_{E}\vert\chi_{s}U(\infty,t)\vert x \rangle\nonumber\\
&\approx&
\sum_{E'}\langle\psi_{E}\vert\chi_{s}\vert\overrightarrow{\psi_{E'}}\rangle\langle\overleftarrow{\psi_{E'}}(t)+R\overrightarrow{\psi_{E'}}(t)\vert x \rangle\nonumber\\
&=&\langle\overleftarrow{\psi_{E}}(t)+R\overrightarrow{\psi_{E}}(t)\vert x \rangle\\
&=&[e^{-\frac{i}{\hbar}p_{s} x}+Re^{\frac{i}{\hbar}p_{s} x}]e^{\frac{i}{\hbar}(E t+eA(t)d)}\nonumber.
\end{eqnarray}
where we use the left and right arrows to represent the backward and forward waves, respectively. Similarly to the approach in Ref.~\cite{Yalunin2011}, the quasi-reflection amplitude $R$ is obtained with the continuity of the wavefunctions. 
\begin{equation}
\label{eq3.3.2}
\begin{split}
R=\frac{e^{-\frac{i}{\hbar}2p_{s}d}(p_{t}+p_{g})(p_{g}-p_{s})}{e^{\frac{i}{\hbar}2p_{g}d}(p_{g}-p_{t})(p_{g}-p_{s})-(p_{t}+p_{g})(p_{g}+p_{s})}\\
+\frac{e^{\frac{i}{\hbar}2(p_{g}-p_{s})d}(p_{t}-p_{g})(p_{g}+p_{s})}{e^{\frac{i}{\hbar}2p_{g}d}(p_{g}-p_{t})(p_{g}-p_{s})-(p_{t}+p_{g})(p_{g}+p_{s})}.
\end{split}
\end{equation}
where $p_{t}=\sqrt{2m[E-V_{0}(x)]},\ (x<0)$ is the momentum in the tip, $p_{s}=\sqrt{2m[E-V_{0}(x)]},\ (x>d)$ is the momentum in the sample, and $p_{g}=\sqrt{2m[E-U_{p}(t)]}$ is a time-dependent momentum in the gap, respectively. Here $U_{p}$ is a ponderomotive energy averaged over the interaction duration  
\begin{eqnarray}\label{eq3.3.3}
U_{p}(t)= \lim_{t_{2}\to\infty} \frac{\int^{t_{2}}_{t}[eA(\tau)]^2\;d\tau\,}{2m(t_{2}-t)}  .
\end{eqnarray}

The quasi-reflection amplitude $R$ is time-dependent and related to the laser field strength. The interpretation is that the backward wave, adiabatically reflected by the sample boundary at time $t$, generates an outward wave with an amplitude of $R$ (see Fig.~\ref{Fig2}(c)). It is assumed that at the moment of reflection, energy is conserved (i.e., no photons are absorbed). Since the wave in the junction oscillates with the laser field, the reflection amplitude $R$ also varies with time. In the absence of the laser field, this amplitude satisfies the static condition, reducing to the exact static reflection amplitude.

Figure~\ref{Fig2}(d) illustrates the tunneling amplitude near zero energy, calculated using three different methods. The result with boundary correction aligns closely with the TDSE, while the uncorrected result exhibits unphysical errors as the energy approaches zero. In Fig.~\ref{Fig1}(c), we present the simulation results with corrections. The spectrum incorporating trajectory selection closely matches the TDSE results, even at energy levels around 15 eV and 0 eV, where the SFA simulation based on the Volkov propagator shows significant deviations.

\section{\label{Sec4}SEMI-CLASSICAL ANALYSIS}
\subsection{\label{sec4.1}Three-step model}

%
% FIGURE 3
%
\begin{figure*}[t!]
\centering
\includegraphics[clip, trim=0cm 0cm 0cm 0cm, width=1\linewidth]{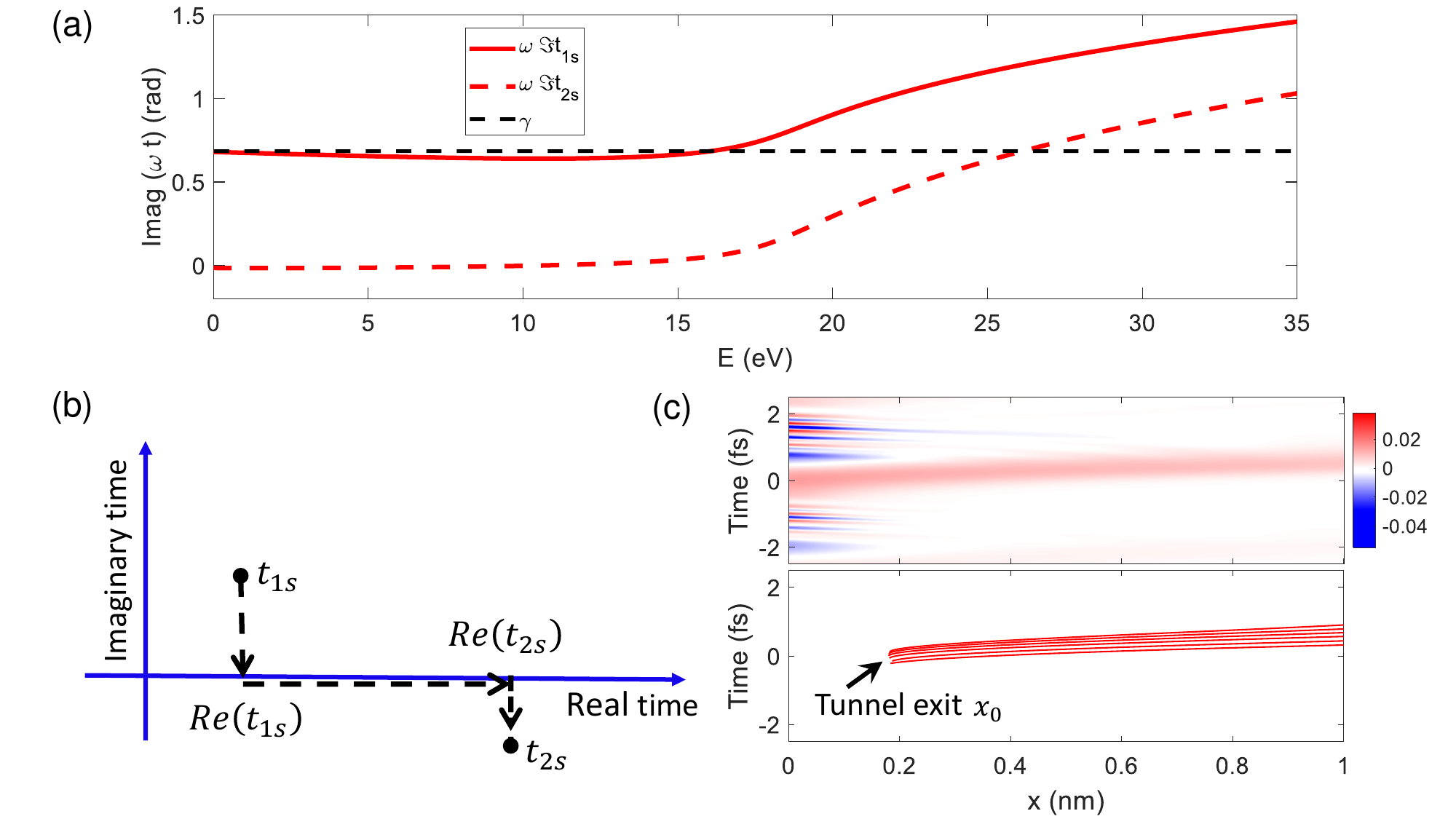}
\caption{Saddle points and integration contour for semi-classical trajectories. (a) Imaginary components of emission time $\Im{t_{\mathrm{1s}}}$ and arrival time $\Im{t_{\mathrm{2s}}}$. Before the cutoff energy, the emission time follows the Keldysh parameter, while the arrival time vanishes. (b) The contour of the displacement. The branch along the imaginary part of $t_{\mathrm{1s}}$ corresponds to ionization, the branch along the real axis represents driven motion, and the imaginary branch of $t_{\mathrm{2s}}$ corresponds to attenuation. (c) Comparisons of TDSE current burst and semi-classical trajectories. Before the tunnel exit is a classically forbidden region. The laser-driven trajectories closely match the TDSE results. In these simulations, the junction width is 1 nm, the field strength is 25 V/nm, and the classical current burst is modeled as a superposition of possible trajectories with final energies ranging from 0 to the cutoff. All other parameters are the same as those used in Fig.~\ref{Fig1}.}
\label{Fig3}
\end{figure*}

So far, we have focused on the derivations using the SFA. Now, we substitute the constrained Van Vleck propagator and the approximated eigenfunction into the formal solution Eq.~\ref{eq2.3.5} and obtain
\begin{eqnarray}\label{eq4.1.1}
M_E&=&\int_{-\infty}^{\infty}\int_{-\infty}^{t_{2}}\sqrt{\frac{i}{8\pi m\hbar^3(t_{2}-t_{1})}}
\eta(t_{1}, t_{2})\nonumber\\
&&\times e^{\frac{i}{\hbar}S(t_{2},t_{1})}\;dt_{1}dt_{2}\,.
\end{eqnarray}
Here $\eta(t_{1}, t_{2})$ is a prefactor coming from transition matrix elements at times $t_{1}$ and $t_{2}$ (see Appendix~\ref{E}). For an initial bound state $E_{0}=-|E_{0}|$ and a final state $E$, the phase factor $S(t_{2},t_{1})$ in the exponent is given by:
\begin{eqnarray}\label{eq4.1.2}
S(t_{2},t_{1})&=&E t_{2}+\frac{{\tilde p}^2}{2m}(t_{2}-t_{1})-\int_{t_{1}}^{t{2}}\frac{e^2A^2(\tau)}{2m} \;d\tau\,\nonumber\\
&&-\int_{t_{1}}^{t{2}}V_{\mathrm{imag}}[x(\tau)]\;d\tau\,+\vert E_{0}\vert t_{1},
\end{eqnarray}
where $\tilde p=\frac{\int_{t_{1}}^{t_{2}}{eA(\tau)}\;d\tau\,+md}{t_{2}-t_{1}}$ is the effective canonical momentum from the Van Vleck propagator. Phase $S(t_{2},t_{1})$ describes a quasiclassical action. The freed electron excited from the initial bound state at time $t_{1}$ travels with the canonical momentum $\tilde p$ in the laser field and obtains an energy $E$ at the final time $t_{2}$. Due to the thin junction, the higher-order rescattering of the wavefunction from the tip boundary has been ignored. 

The exponential factor in Eq.~\ref{eq4.1.1} varies much more rapidly than the other factors, justifying the use of the saddle point method~\cite{Lewenstein1994,Ivanov2005,Pedatzur2015}. The stationary points of this integration are the extrema points of the exponent's phase,
\begin{equation}\label{eq4.1.3}
\nabla_{t_{1},t_{2}}S(t_{2},t_{1})=0.
\end{equation}
Combined with the definition of the effective canonical momentum,
we obtain the following three saddle-point equations:
\begin{eqnarray}
\frac{\left[\tilde p_\mathrm{(s)}-eA(t_{\mathrm{1s}})\right]^2}{2m}-\overline{|V_{\mathrm{imag}}|}=-\left\vert E_0\right\vert,\label{eq4.1.4}
\\
\int_{t_\mathrm{1s}}^{t_\mathrm{2s}}\frac{\left[\tilde p_\mathrm{(s)}-eA(\tau)\right]}{m}\;d\tau\,=d,\label{eq4.1.5}
\\
\frac{\left[\tilde p_\mathrm{(s)}-eA(t_\mathrm{2s})\right]^2}{2m}-\overline{|V_{\mathrm{imag}}|}=E.\label{eq4.1.6}
\end{eqnarray}
Here, the subscript (s) indicates that these values correspond to mathematical saddle points rather than physical quantities. $\overline{|V_{\mathrm{imag}}|}$ represents the averaged absolute value of the image potential defined in Fig.~\ref{Fig1}(b). Similarly to the three-step model for HHG~\cite{Lewenstein1994,Corkum2007} and high-order ATI~\cite{Salieres2001,Milosevic2006}, the three equations above provide a three-step framework for describing electron transport in the laser-driven MIM structure.
(1) At the time of emission, the bound electron is freed from the tip by the laser field. Equation~\ref{eq4.1.4} describes the energy relationship at this instant. The image potential lowers the barrier, enhancing the emission of the electron. 
(2) The freed electron is accelerated by the laser field, moving from the tip to the sample. Equation~\ref{eq4.1.5} describes its dynamics. Unlike the hallmark recollision processes of attosecond science, the electron does not return to its origin. The force exerted by the image potential is disregarded.
(3) Transmission into the sample: The energy accumulated during the second step is transferred to the sample, as described by Eq.~\ref{eq4.1.6}. It should be stressed that the initial energy is negative, leading to complex values of $t_{1s}$, $\tilde p_{(s)}$, and $t_{2s}$. Figure~\ref{Fig3}(a) shows the imaginary parts of the emission time $t_{1s}$ and the arrival time $t_{2s}$. Before the cutoff energy, the emission time takes on an imaginary value related to the Keldysh parameter $\gamma={\omega\sqrt{2m|E_{0}|_{\mathrm{eff}}}/{|e|F}}$, where $F$ is the laser field strength, $|E_{0}|_{\mathrm{eff}}=|E_{0}|-\overline{|V_{\mathrm{imag}}|}$ is the effective binding energy, and $\omega$ is central frequency. The imaginary part of the emission time corresponds to purely quantum processes, such as photon absorption and tunneling through the barrier. In contrast, the arrival time $t_{2s}$ is almost exclusively real, which causes the canonical momentum $\tilde p_{(s)}$ to be almost completely real as well. For $\gamma\ll 1$, the vanishing imaginary part of $\tilde p_{(s)}$ results in 

\begin{equation}\label{eq4.1.7}
\frac{\int_{\Re{t_{1s}}}^{\Re{t_{2s}}}{eA(\tau)}\;d\tau\,+m(d-x_{0})}{\Re{t_{2s}}-\Re{t_{1s}}}-eA(\Re{t_{1s}})=0,
\end{equation}
where notation $\Re$ denotes the real parts and $x_{0}=\int_{t_{1s}}^{\Re{t_{1s}}}{\frac{\tilde p-eA(\tau)}{m}}\;d\tau\,$ is the tunnel exit. This formula can be interpreted as indicating that the freed electron has no kinetic energy at the tunnel exit, which is consistent with the adiabatic tunneling picture~\cite{Corkum1993,Paulus1994a,Zheltikov2016}. The imaginary part of the arrival time appears only when the final energy $E$ reaches the cutoff. Therefore, the imaginary component of $t_{2s}$ corresponds to attenuation. 

In the adiabatic regime ($\gamma\ll1$), the saddle-point solutions also offer a semi-classical trajectory scenario. In Newtonian mechanics, the displacement of a classical point-like electron in the laser field is 
\begin{equation}\label{eq4.1.8}
{\cal D}(t)=\int_{t_\mathrm{1s}}^{t}\frac{\left[\tilde p_{(\mathrm{s})}-eA(\tau)\right]}{m}\;d\tau\,.
\end{equation}
Since the times are complex, the integral must be taken over a complex contour (Fig.~\ref{Fig3}(b)). Each branch corresponds to a specific step in the three-step model. The trajectory begins with a tunnel exit, after which it follows a path that leads to the sample (Fig.~\ref{Fig3}(c)). 

%
% FIGURE 4
%
\begin{figure*}[t!]
\centering
\includegraphics[clip, trim=0cm 6cm 0cm 3cm, width=1\linewidth]{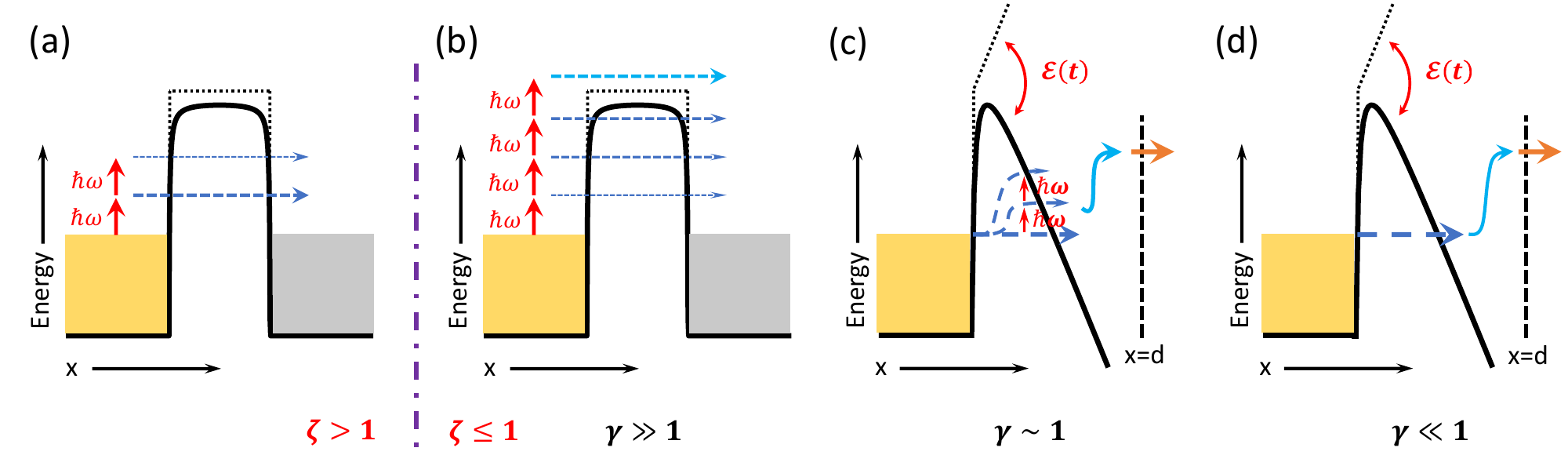}
\caption{Photoemission in different regimes. (a) Photon excitation under the barrier. The electron in this regime is excited to a higher eigenstate below the threshold energy. This regime is associated with $\zeta>1$. (b) Multiphoton regime. The bound electron reaches the threshold energy by absorbing multiple photons. The parameter $\zeta$ becomes smaller than 1. (c) Non-adiabatic regime. The bound electron absorbs photons while tunneling through the barrier and is subsequently accelerated by the laser field after exiting the barrier. (d) Laser-induced tunneling (adiabatic) regime. The electron tunnels through the barrier, is driven by the laser field, and ultimately transmits into the sample.}
\label{Fig4}
\end{figure*}

\subsection{\label{sec4.2}Classification of laser-driven processes in the MIM junction}

Due to the narrow width of the junction and the strong laser field, electron emission is highly localized near the crest of the waveform, with trajectories traversing the junction within half an optical cycle, as illustrated in Fig.~\ref{Fig3}(c). Consequently, in this section, we analyze the photoemission mechanisms within a single optical cycle. To simplify the few-cycle laser pulse, we only focus on the central laser frequency within a short time domain and set the static bias voltage $U_{\mathrm{s}}=0$:
\begin{equation}\label{eq4.2.1}
A(t)=\frac{F}{\omega}\sin{(\omega t)}.
\end{equation}
Substituting this formula and $\Im{t_{2s}}=0$ into Eq.~\ref{eq4.1.4} and Eq.~\ref{eq4.1.6}, and subsequently subtracting the equations and extracting the imaginary terms, we obtain $\sinh(\omega\Im{t_{1s}})\sim{\omega\sqrt{2m|E_{0}|_{\mathrm{eff}}}}/{|e|F}=\gamma$, with
\begin{equation}\label{eq4.2.2}
\Im{t_{1s}}=\frac{1}{\omega}\operatorname{arsinh}(\gamma)=\frac{1}{\omega}\ln(\gamma+\sqrt{1+\gamma^2}).
\end{equation}
where $\operatorname{arsinh}[...]$ is the inverse hyperbolic sine function. Eq.~\ref{eq4.2.2} is a well-known solution in the strong-field ionization of atomic systems~\cite{Keldysh1965,Zheltikov2016,Popruzhenko2014}. It characterizes the regime transition in terms of the Keldysh parameter $\gamma$. Under saddle-point analysis, the current emitted from the tip is proportional to the exponential of the imaginary component of the action $S$~\cite{Milosevic2006,Keldysh1965,Ivanov2005,Yalunin2011}:
\begin{equation}\label{eq4.2.3}
{\cal J}\propto e^{-\frac{2}{\hbar}\Im{[S(t_{2s},t_{1s})]}}.
\end{equation}

%
% FIGURE 5
%
\begin{figure*}[t!]
\centering
\includegraphics[clip, trim=0.5cm 3.5cm 0cm 2cm, width=1\linewidth]{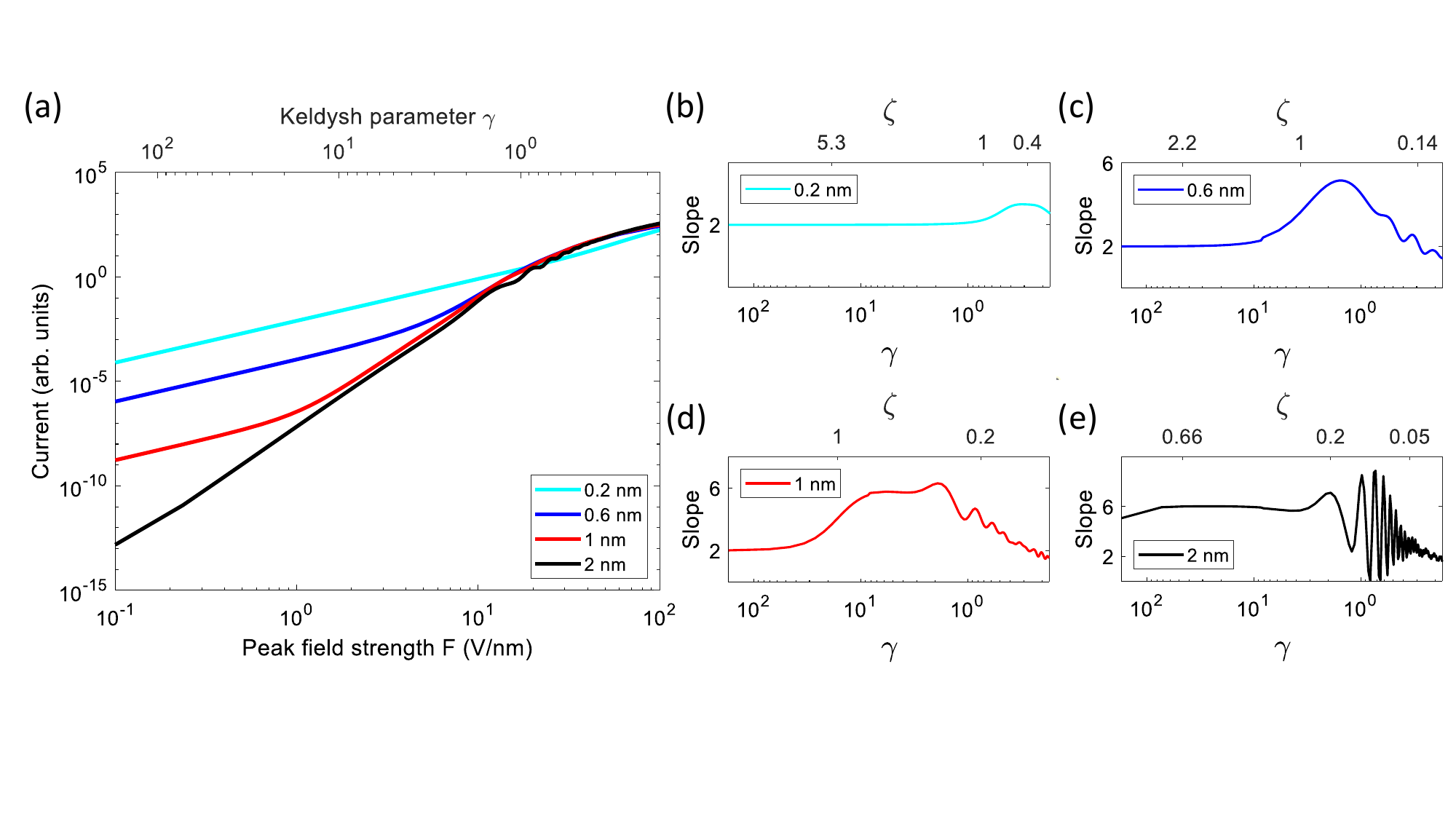}
\caption{(a) TDSE simulations of the current as a function of field strength for different junction widths. The plot is presented in log-log scale, with the corresponding Keldysh parameters listed above. (b)-(d) The slopes of the currents as a function of field strength in (a). The photoemission can transition to the strong-field regime only when $\zeta<1$. The simulation parameters are the same as those used in Fig.~\ref{Fig1}.}
\label{Fig5}
\end{figure*}

To classify the photoemission, here we define a new parameter $\zeta$, given by
\begin{equation}\label{eq4.2.4}
{\zeta}=\frac{\gamma\ln(\gamma+\sqrt{1+\gamma^2})}{2\delta},
\end{equation}
where $\delta=\frac{\omega^2 m d}{|e|F}$ is a length parameter introduced for spatially confined nanostructures by Herink \textit{et al.}~\cite{Herink2012}. The properties, derivation, and interpretation of $\zeta$ will be discussed in detail in the following two sections. After some algebra (see Appendix~\ref{F}) we obtain the following results:

(a) For weak interactions with $\zeta>1$, the photocurrent is given by
\begin{equation}\label{eq4.2.5}
{\cal J}\propto \exp\bigg[{-\frac{2}{\hbar}d\sqrt{2m(\ |E_{0}|_{\mathrm{eff}}-n\hbar\omega\ )}}\bigg],
\end{equation}
where $n=\frac{|e|Fd}{2\hbar\omega}$ represents the effective photon number. This expression shows an exponential decay with respect to the barrier width $d$, which is a characteristic feature of tunneling under a rectangular barrier (Fig.~\ref{Fig4}(a)). In the absence of the laser field, the formula reduces to $\exp[{-\frac{2}{\hbar}d\sqrt{2m |E_{0}|_{\mathrm{eff}}}}]$, corresponding to static tunneling in the nanojunction. This result confirms that our strong-field theory is applicable even in the weak perturbation regime.

(b) As the interaction increases to $\zeta\leq1$, the photoemission can be characterized by the Keldysh parameter $\gamma$. For $\gamma\gg1$, we find $\Im{t_{1s}}=\ln(2\gamma)/\omega$, corresponding to the multiphoton regime (Fig.~\ref{Fig4}(b)). In this regime, photon absorption reaches the threshold of the effective work function, and the photocurrent is given by:
\begin{equation}\label{eq4.2.6}
{\cal J}\propto (2\gamma)^{-2\frac{|E_{0}|_{\mathrm{eff}}}{\hbar\omega}}.
\end{equation}
The current is dominated by the laser.

(c) For non-adiabatic conditions ($\gamma\sim1$), performing a Taylor expansion of $\Im{t_{1s}}$ at $\gamma=1$, the photocurrent becomes
\begin{eqnarray}\label{eq4.2.7}
{\cal J}&&\propto [(\gamma+\sqrt{1+\gamma^2})]^{-2\frac{|E_{0}|_{\mathrm{eff}}}{\hbar\omega}}\nonumber\\
&&\ \ \times\exp[{-\frac{(2m|E_{0}|_{\mathrm{eff}})^{\frac{3}{2}}}{3\hbar m|e|F}}].
\end{eqnarray}
This formula includes both a photon absorption term and an exponential tunneling term (Fig.~\ref{Fig4}(c)). In this regime, the barrier is not static, and electrons absorb photon energy while tunneling through the laser-suppressed barrier, resulting in an energy shift at the tunnel exit. This result aligns with the findings in Ref.~\cite{Ivanov2005}.

(d) For the adiabatic condition ($\gamma\ll1$), we find $\Im{t_{1s}=\gamma/\omega}$. Here, the photoemission occurs in the laser-induced tunneling regime (Fig.~\ref{Fig4}(d)):
\begin{equation}\label{eq4.2.8}
{\cal J}\propto \exp\bigg[{-\frac{2(2m|E_{0}|_{\mathrm{eff}})^{\frac{3}{2}}}{3\hbar m|e|F}}\bigg],
\end{equation}
This is the well-known Fowler-Nordheim field emission formula for a stationary field without image potential corrections~\cite{Forbes2003,Fowler1928}. In this case, the tunnel exit depends on the laser field instead of the junction width $d$.

These results indicate that for $\zeta>1$, the current is primarily governed by the static junction, whereas for $\zeta\leq1$, it is dominated by the laser field. In the case of $\zeta>1$, photon excitation is under the junction barrier. In contrast, for $\zeta\leq1$, the electron is first ionized from its initial state to a free state (spatially confined Volkov state) in the vacuum gap and then driven across the junction.

In Fig.~\ref{Fig5}, we simulate the laser-induced tunneling current as a function of field strength for different junction widths. The slopes for $\gamma>1$ correspond to twice the effective number of absorbed photons. For $\gamma>1$ with $\zeta<1$, currents for 1 nm and 2 nm show 2.85 and 3 effective photon absorption, respectively (Fig.~\ref{Fig5}(d) and (e)). For a 0.6 nm gap with $\zeta<1$, the current is dominated by an effective 2.5-photon absorption process for $\gamma>1$ (Fig.~\ref{Fig5}(c))). These values are slightly below the threshold of $|E_{0}|/\hbar\omega=3.34$ due to the contribution of the image effect. For an extremely thin junction of 0.2 nm, as well as for other junction widths with $\zeta>1$, the multiphoton process does not occur at any field strength. Instead, single-photon absorption dominates. This behavior arises from the direct constraint imposed by the parameter $\zeta$, which permits only a weak excitation below the junction, as described by Eq.~\ref{eq4.2.5}. As the field strength increases and the Keldysh parameter drops below 1 ($\gamma<1$), the curves begin to merge for all junction widths, except the curve for the 0.2 nm junction. This indicates the transition to the laser-induced tunneling regime. In this regime, the laser field completely dominates the process and the suppressed tunneling barrier is thinner than the junction width (Fig.~\ref{Fig4}(d)). The curves for different junction widths converge to the same tunneling current since they all have the same tunnel exit, $x_{0}$. The curve for the 0.2 nm junction, however, does not make the transition into the tunneling regime until $\zeta<1$. For a field strength of 25 V/nm, the tunnel exit is at a position of 0.2 nm, which is comparable to the width of the junction. The thinness of the junction enhances static tunneling, making it consistently stronger than other nonlinear processes.

The curves for junction widths larger than 0.6 nm exhibit rapid oscillations around $\gamma=1$, a phenomenon attributed to channel closings caused by the increase in ponderomotive energy (Fig.~\ref{Fig5}(c), (d), and (e)). During the optical cycle, the electron quivers and possesses ponderomotive energy. However, the spatial confinement of the nanojunction limits this quiver motion. Sufficient ponderomotive energy can only accumulate if the junction width exceeds the quiver amplitude. This condition is characterized by the length parameter $\delta$~\cite{Herink2012}. For MIM junctions, the parameter $\delta$ is defined as the ratio of the junction width to the quiver amplitude, and sufficient ponderomotive energy accumulation occurs when $\delta>1$. In our simulations, at $\gamma=1$, the 0.6 nm, 1 nm, and 2 nm junctions yield $\delta$ values of 1.04, 1.72, and 3.45, respectively, satisfying the condition for channel closing. Conversely, junction widths smaller than 0.6 nm do not meet this criterion and therefore cannot support sufficient channel closing.

To compare with our SFA method, we also present the same calculations for all three approaches in Appendix~\ref{G}. The SFA with all corrections closely matches the TDSE results. In contrast, the Volkov SFA approach fails to filter out unphysical trajectories and significantly amplifies the ponderomotive oscillations around $\gamma=1$, which should only occur for large gap widths. Channel closings in the TDSE calculations and the SFA calculations including all corrections are observed only when the gap width $d\geq0.6\ \mathrm{nm}$.

\subsection{\label{sec4.3}Cutoff energy}

\begin{figure} %0.6\columnwidth
\includegraphics[clip, trim=10.5cm 0cm 10cm 1cm,width=0.9\columnwidth]{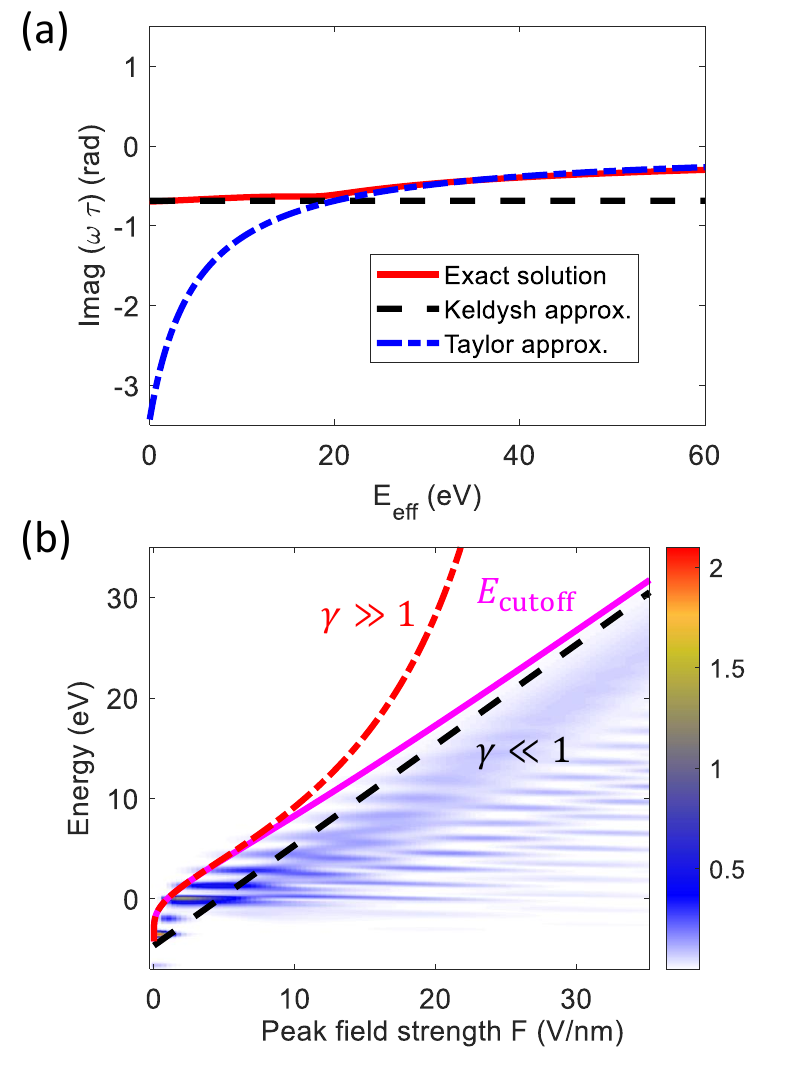}
\caption{Cutoff energy of the tunneling spectrum. (a) The two approximated solutions for the duration $\tau$ are compared with the numerical result. The intersection point is defined as the cutoff. (b) The cutoff energy is compared with the TDSE spectrum. Approximations for the adiabatic and multiphoton regimes are included for comparison. In these simulations, the junction width is set to 1 nm. Panel (a) is calculated with a field strength of 25 V/nm, while the TDSE simulations in panel (b) use a 6 fs, 830 nm pulse. Other parameters match those used in Fig.~\ref{Fig1}. }
\label{Fig6}
\end{figure}

As discussed in Sec.~\ref{sec4.1}, the imaginary component of the arrival time $t_{2s}$ remains zero until the energy $E$ reaches the cutoff. In this section, we aim to derive a cutoff law that accommodates the photoemission types described above. The cutoff energy represents the maximum energy the electron can classically acquire during these processes. Based on our preceding derivations in Sec.~\ref{sec4.2}, we obtained the imaginary components of the duration $\tau=t_{2s}-t_{1s}$ is $-\Im{t_{\mathrm{1s}}}$ until the final energy $E$ reaches the cutoff. After the cutoff, the duration decreases strongly. For this case, $\tau$ can be expressed using a Taylor expansion,
\begin{equation}\label{eq4.3.1}
\tau=\frac{2md}{\sqrt{2mE_{\mathrm{eff}}}+i\sqrt{2m|E_{0}|_{\mathrm{eff}}}},
\end{equation}
where $E_{\mathrm{eff}}=E+\overline{|V_{\mathrm{imag}}|}$. Fig.~\ref{Fig6}(a) illustrates the agreement between the two approximated solutions and the numerical results. Each solution aligns well with the exact solution in its respective spectral region. Notably, the duration before the cutoff maintains a constant imaginary value, while Eq.~\ref{eq4.3.1} exhibits a varying imaginary value. In the latter case, the change in the imaginary part leads to significant attenuation of the tunneling amplitude. Consequently, we define the intersection of these two approximated solutions as the cutoff energy. By equating the imaginary parts of these solutions, we derive
\begin{equation}\label{eq4.3.2}
E_\mathrm{cutoff}=\frac{|E_{0}|_{\mathrm{eff}}}{\zeta}-|E_{0}|,
\end{equation}
where $\zeta$ is defined in Eq.~\ref{eq4.2.4}. With this relationship, we can examine how the cutoff energy varies under different conditions.

%
% FIGURE 6
%

(a) For the adiabatic case ($\gamma\ll 1$), the barrier can be treated as static while the electron moves through it (Fig.~\ref{Fig4}(d)). Eq.~\ref{eq4.3.2} simplifies to 
\begin{equation}\label{eq4.3.3}
E_\mathrm{cutoff}=|e|Fd-|E_{0}|.
\end{equation}
This cutoff corresponds to the energy shift in a completely static homogeneous field, which is also the maximum energy acquired in the ultrafast STM~\cite{Ma2024}.

(b) For the multiphoton regime ($\gamma\gg 1$), the electron does not undergo tunneling ionization but can only be ionized through photon absorption (Fig.~\ref{Fig4}(b)). In this case, the cutoff energy is reduced to:
\begin{equation}\label{eq4.3.4}
E_\mathrm{cutoff}=\bigg[\frac{d\sqrt{2m|E_0|_{\mathrm{eff}}}}{\hbar\ln(2\gamma)}\bigg]\hbar\omega-|E_{0}|.
\end{equation}
The factor in the square bracket is dimensionless and limits the effective maximum number of absorbed photon. Furthermore, as $\gamma$ approaches infinity ($F=0$), which can be interpreted as the laser-free case, the cutoff energy is $E_\mathrm{cutoff}=-|E_0|=E_0$, equivalent to the energy relation in stationary tunneling.

(c) For the non-adiabatic case ($\gamma\sim 1$), photoemission is a mixture of photon absorption and tunneling ionization (Fig.~\ref{Fig4}(c)). The cutoff energy will be slightly higher than the adiabatic cutoff energy due to photon absorption. For instance, we find $E_\mathrm{cutoff}=1.1346|e|Fd-|E_{0}|$ for $\gamma=1$. To clearly illustrate this phenomenon, we can perform a Taylor expansion around $\gamma=1$. Therefore, Eq.~\ref{eq4.3.2} can be rewritten as a series
\begin{equation}\label{eq4.3.5}
E_\mathrm{cutoff}=a_{0}|e|Fd-|E_{0}| +\bigg[\frac{d\sqrt{2m|E_0|_{\mathrm{eff}}}}{\hbar}(\sum_{n=1}^{\infty}a_{n}\gamma^{n})\bigg]\hbar\omega.
\end{equation}
where the coefficient $a_{n}$ comes from the Taylor expansion of $\gamma/\ln(\gamma+\sqrt{\gamma^2+1})$, with the zeroth order $a_{0}\sim1$. This indicates that the tunneling process is still dominant. The result is consistent with the conclusion in Sec. ~\ref{sec4.2}.

Fig.~\ref{Fig6}(b) shows the cutoff plot of Eq.~\ref{eq4.3.2} compared to the TDSE tunneling spectrum, demonstrating a match across all regimes. For comparison, the adiabatic approximation (Eq.~\ref{eq4.3.3}) and multiphoton approximation (Eq.~\ref{eq4.3.4}) are also shown.

\subsection{\label{sec4.4}Discussion of the new parameter $\zeta$}

In strong-field theories, the Keldysh parameter $\gamma$ plays an important role in characterizing the photoemission mechanisms. However, in the preceding two sections, we frequently encountered a new dimensionless parameter $\zeta$. This parameter is a function of the field strength and junction width, and it vanishes as the width increases. Our results in Fig.~\ref{Fig5} suggest that the Keldysh parameter alone is insufficient for describing interactions and regime transitions in nanojunctions, and a new parameter is required.

%
% FIGURE 7
%
\begin{figure} %0.6\columnwidth
\includegraphics[clip, trim=11cm 5cm 9cm 2.5cm, width=0.9\columnwidth]{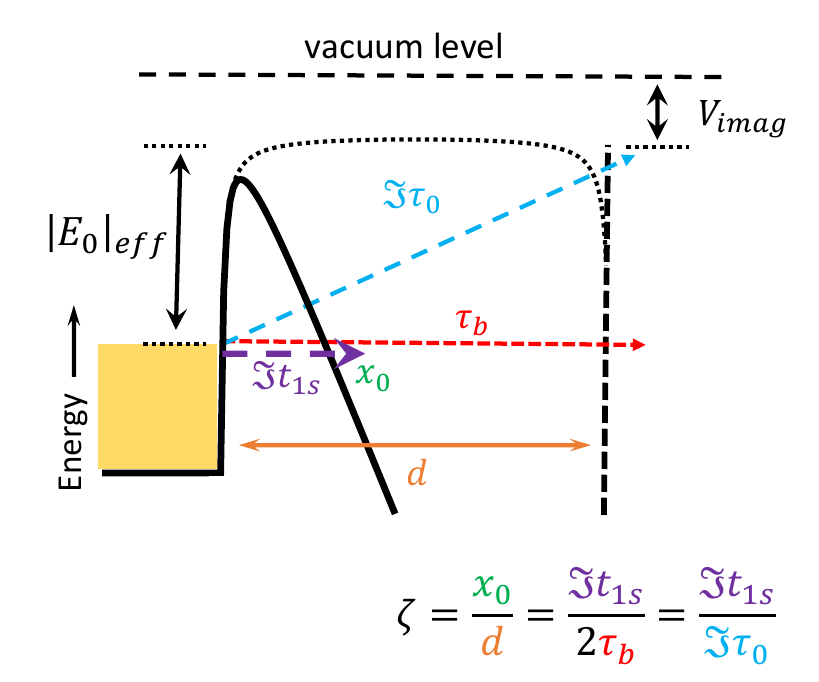}
\caption{Scheme for different interpretations of $\zeta$. The figure is presented exclusively in the adiabatic regime ($\gamma \ll 1$). Key parameters include the junction width $d$, tunnel exit $x_{0}$, intrinsic tunneling times $\tau_{\mathrm{b}}$, and $\Im{\tau_{0}}$.}
\label{Fig7}
\end{figure}

The issue with the Keldysh parameter arises from the term ``strong-field''. While a strong field is used to stimulate the system, the interaction is ultimately governed by the dipole potential, which depends on both the field strength \textit{and} the displacement~\cite{Amini2019}. In a system without spatial constraints, a strong field strength is equivalent to a strong interaction. However, the situation differs for a nanojunction with confined boundaries, where a strong field coupled with a small junction may result in a weak dipole potential. The Keldysh parameter, which only accounts for the field strength, cannot fully characterize the interactions in MIM junctions.

The new parameter, $\zeta$, is defined in terms of the Keldysh parameter $\gamma$ and the length parameter $\delta=\frac{\omega^2 md}{|e|F}$. This parameter addresses the limitations of the Keldysh parameter in our context. According to Zheltikov's interpretation, the Keldysh parameter describes the time required for an electron to acquire ponderomotive energy equal to the ionization potential~\cite{Zheltikov2016}. However, as discussed earlier, in a nanojunction, significant ponderomotive energy does not accumulate unless $\delta>1$. Consequently, $\delta$ determines whether the Keldysh parameter is applicable in such scenarios. By combining $\gamma$ and $\delta$, the parameter $\zeta$ bridges the gap where $\gamma$ alone falls short. 

However, the above definition is based on the scenario of zero static bias, with only the laser field present. Next, we develop a more general definition of $\zeta$ derived from its physical properties. The Keldysh parameter $\gamma$ can be expressed as $\gamma=4\pi\tau_{b}/T$, where $\tau_{b}=dm/\sqrt{2m|E_{0}|_{\mathrm{eff}}}$ is the characteristic time that the electron spends traversing the junction barrier and $T=2\pi/\omega$ is the central period of the laser field~\cite{Zheltikov2016,Keldysh1965}. This formulation provides a temporal interpretation of $\gamma$, linking it to the interplay between the intrinsic barrier traversal time and the duration of the laser cycle. In contrast, the parameter $\zeta$ can be expressed as
\begin{equation}\label{eq4.4.1}
\zeta=\frac{\sqrt{2m|E_{0}|_{\mathrm{eff}}}}{2md}\frac{\ln(\gamma+\sqrt{1+\gamma^2})}{\omega}=\frac{\Im t_{\mathrm{1s}}}{2\tau_{b}}=\frac{\Im t_{\mathrm{1s}}}{\Im{\tau_{0}}},
\end{equation}
where $\Im t_{\mathrm{1s}}$ is the strong-field ionization time defined in Eq.~\ref{eq4.2.2} and $\Im{\tau_{0}}$ is the imaginary component of $\tau=2md/(\sqrt{2m|E_{0}|_{\mathrm{eff}}}+\sqrt{2m|E|_{\mathrm{eff}}})$ at the threshold $|E|_{\mathrm{eff}}=0$, representing the shortest intrinsic tunneling time (see Fig.~\ref{Fig7}). Without bias, Eq.~\ref{eq4.4.1} is equivalent to Eq.~\ref{eq4.2.4}, but it provides an interpretation of $\zeta$ as the ratio of the strong-field ionization time to the intrinsic tunneling time. For a strong process ($\zeta<1$), the laser ionization mechanism, via tunneling or multiphoton absorption, must act faster than the intrinsic tunneling time. Otherwise, the electron would complete junction tunneling before the strong-field ionization, rendering the energy transfer ineffective. This interpretation also applies to Fig.~\ref{Fig6}(a). The two approximated solutions represent two different photoemission processes. The real ionization (numerical result) selects only the fastest pathway, with the cutoff serving as the boundary. The definition $\zeta=\frac{\Im{t_{\mathrm{1s}}}}{\Im{\tau_0}}$ is based solely on time parameters, making it easily extendable to other conditions, such as photoemission with bias.

In the following, we will illustrate how $\zeta$ influences the interaction from both adiabatic and energetic perspectives. The length parameter $\delta$ is an adiabatic parameter defined as a ratio of the junction width and the quiver amplitude~\cite{Herink2012}.  Similar to this definition, in the adiabatic regime ($\gamma\ll1$), $\zeta$ can be expressed as
\begin{equation}\label{eq4.4.2}
\zeta=\frac{\gamma^2}{2\delta}=\frac{\frac{|E_{0}|_{\mathrm{eff}}}{|e|F}}{d}=\frac{x_{0}}{d}.
\end{equation}
where $x_{0}=\frac{|E_{0}|_{\mathrm{eff}}}{|e|F}$ represents the tunnel exit. Consequently, $\zeta$ is the ratio of the tunnel exit to the junction width (see Fig.~\ref{Fig7}). The value of $\zeta$ provides insight into the dynamics of tunneling. 

(a) $\zeta\ll1$: The laser field is extremely strong, causing the tunnel exit to approach the tip. This results in efficient tunneling, as the electron experiences a suppressed barrier. 

(b) $\zeta\sim1$: The tunnel exit is located near the sample boundary, leaving insufficient space for the emitted electron to accelerate significantly. This condition limits the energy absorption.

(c) $\zeta>1$: The laser-induced tunneling vanishes as the tunnel exit lies beyond the junction width.

Considering the definition of the cutoff in Eq.~\ref{eq4.3.2}, the parameter $\zeta$ can also be written as
\begin{equation}\label{eq4.4.3}
\zeta=\frac{\gamma\ln[\gamma+\sqrt{1+\gamma^2}]}{2\delta}=\frac{|E_{0}|_{\mathrm{eff}}}{E_{\mathrm{cutoff}}-|E_{0}|}.
\end{equation}
The parameter $\zeta$ represents the energy capacity ratio, comparing the system's initial energy to the maximum energy absorbed from the laser. For a small $\zeta$, the energy capacity is large, allowing the MIM junction to exhibit laser-dominated photoemission processes. Conversely, for a large $\zeta$, the energy capacity is small, limiting the transfer of laser energy to the photoelectrons, even if the laser field is strong. For a given nanojunction, $\zeta=1$ indicates that the maximum energy only reaches the threshold of the junction. When $\zeta>1$, only photonic excitation below the work function is possible. In contrast, $\zeta<1$ allows for multiphoton photoemission, above-threshold photoemission, and laser-induced tunneling emission.

Our strong-field theory is applicable only to nanojunctions. For larger junctions, the cutoff energy $E_{\mathrm{cutoff}}$ of the direct electron will eventually approach $2U_{p}$~\cite{Herink2012}. In such cases, we find $\zeta=\frac{\gamma^2}{1-\gamma^2}$ that is related solely to the Keldysh parameter $\gamma$. Therefore, the Keldysh parameter, which is independent of the junction width, can sufficiently characterize photoemission in these cases.

\subsection{\label{sec4.5}Carrier-envelope phase modulation}

Electron emission in the laser-assisted nanojunction occurs on the attosecond timescale, enabling ultrafast coherent control. In symmetric nanojunctions (the same materials, with no bias), CEP modulation enables control over the direction of the current and electron trajectories by adjusting the CEP of the laser pulse. In this section, we discuss CEP modulation of the tunneling current. CEP control is highly sensitive to the duration of the laser pulse. For long-duration pulses containing multiple cycles, current bursts in each cycle tend to cancel each other out. In contrast, for few-cycle pulses, variations in the CEP can induce sufficient symmetry breaking. This technique is both simple and efficient, preserving quantum coherence. Experimental verification and theoretical discussions of this approach have also been conducted in the context of attosecond STM~\cite{Garg2020,Ma2024}.

%
% FIGURE 8
%
\begin{figure*}[t!]
\centering
\includegraphics[clip, trim=2.3cm 5cm 3cm 3cm, width=1\linewidth]{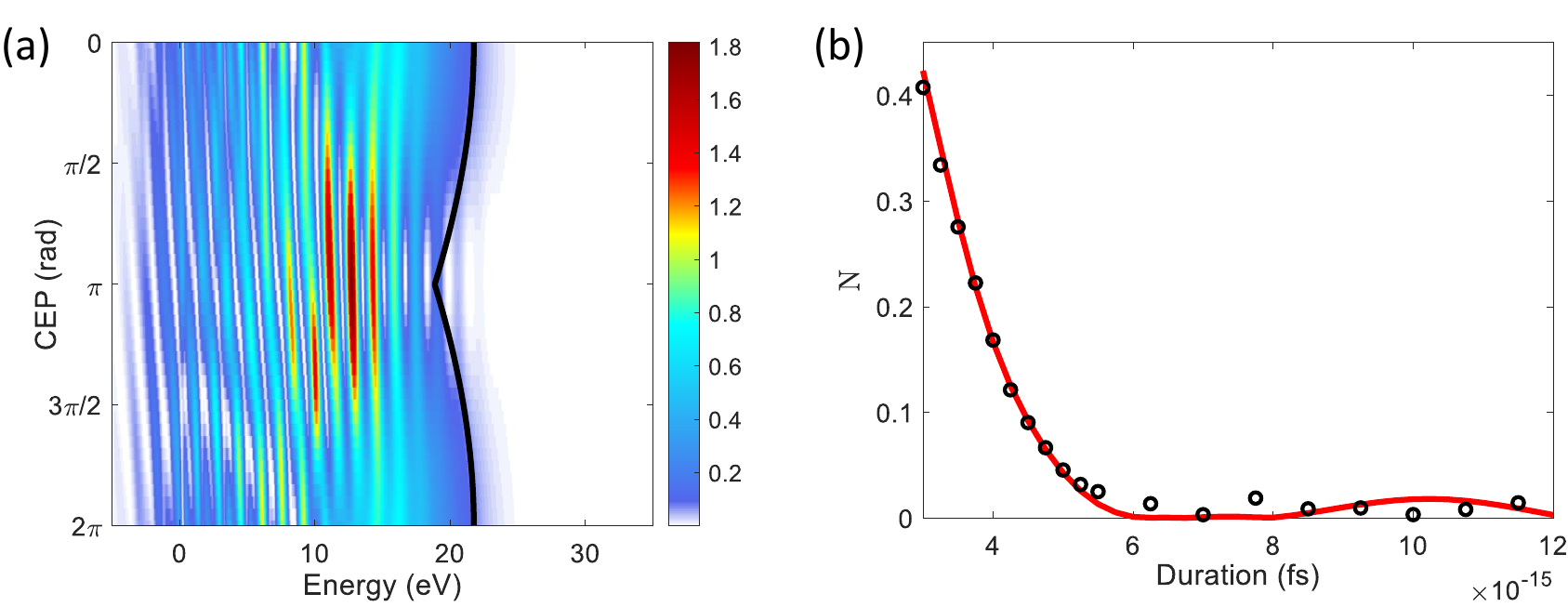}
\caption{CEP modulation in the tunneling spectrum and the effect of pulse duration. (a) The tunneling spectrum (color map) calculated using the TDSE. The junction width is 1 nm, the field strength is 25 V/nm, and the laser pulse duration is 2.8 fs (central wavelength: 830 nm). All other parameters are the same as in Fig.~\ref{Fig1}(c). The black solid curve represents the theoretically predicted cutoff energy, calculated using the effective field strength. (b) The visibility $\cal{N}$ of the CEP current modulation as a function of laser pulse duration. The CEP modulation rapidly diminishes and becomes negligible beyond 6 fs (2 cycles). The red solid curve is the TDSE result and the circles show the SFA result.}
\label{Fig8}
\end{figure*}

We first examine the effect of the CEP on the tunneling spectrum. Figure~\ref{Fig8}(a) illustrates how the tunneling spectrum varies with CEP. A 6-fs laser pulse with three carrier cycles is used for this analysis. The bandwidth of the electron spectrum oscillates periodically with the CEP as the CEP influences the effective maximum field strength, which determines the cutoff energy. The strongest and weakest emissions occur at CEP values of 0 and $\pi$, respectively.

To evaluate the ability of current rectification control, we propose testing it based on the independent current assumption. A ratio parameter ${\cal N}$ is defined to quantify the visibility of CEP modulation:
\begin{equation}\label{eq4.5.1}
{\cal N}=\bigg\vert\frac{I_{\mathrm{t}}(0)-I_{\mathrm{s}}(\pi)}{I_{\mathrm{t}}(0)+I_{\mathrm{s}}(\pi)}\bigg\vert.
\end{equation}
where $I_{\mathrm{t}}$ and $I_{\mathrm{s}}$ represent the tunneling currents emitted from the tip and sample, respectively. Parameters 0 and $\pi$ denote the CEP values. A high ${\cal N}$ indicates strong CEP modulation, whereas ${\cal N}=0$ signifies that the currents cancel out and the CEP effect is averaged out. Fig.~\ref{Fig8}(b) illustrates the dependence of the ratio ${\cal N}$ on the duration of the pulse. The results show that the CEP modulation is sufficiently strong for a single-cycle pulse (3 fs) but vanishes above a two-cycle pulse (6 fs). This finding suggests a direct yet challenging approach to achieving CEP modulation: employing near-single-cycle laser techniques~\cite{Wirth2011,Rybka2016,Ludwig2020}. Such techniques are essential for attosecond observations in MIM nanojunctions.

\section{\label{Sec5}CONCLUSION}

In this paper, we have introduced a refined time-dependent strong-field theory for attosecond electron transport in metal-insulator-metal nanostructures. Unlike our conventional SFA theory formulated for ultrafast STM~\cite{Ma2024}, our approach incorporates both the image potential and boundary effects. To this end. we employ the Van Vleck propagator, rather than the commonly used Volkov propagator, to filter out unphysical trajectories. This propagator further incorporates the image potential into the phase, providing insight into how this weak effect influences the time evolution of the tunneling electron wavepacket. The results obtained with our improved SFA theory demonstrate excellent agreement with numerical TDSE calculations.

Our theory accurately describes processes across different regimes, including one-photon excitations. As a theory for nanojunctions, it takes the form of Bardeen's tunneling matrix element, a standard time-independent theory for static tunneling transmission in MIM devices. Our derivation shows that Bardeen's tunneling matrix can be extended to time-dependent conditions without relying on any approximations (Eq.~\ref{eq2.3.5}). 

We derive saddle-point equations by analyzing the quasi-classical action. These three equations govern energy conservation in the emission and transmission processes, as well as the motion of the electron in the nanojunction. Under strong fields with $\gamma<1$, we identify a three-step picture for the tunneling regime. This picture illustrates how the laser field frees an electron, drives it to the other metal contact, and induces charge transport in the nanojunction. In contrast to the well-known three-step model in HHG, our model does not return the moving electron to its parent ion but instead transports it to the other contact. To study electron transport in the multiphoton and non-adiabatic regimes, we derive an energy cutoff law applicable across all regimes. This law limits the extent of the tunneling spectra and reveals the underlying quantum processes. Using the three-step model and the cutoff law, we successfully address several key phenomena: photon absorption in the multiphoton regime, initial energy shifts in the non-adiabatic regime, the merging behavior of current-field scaling curves, adiabatic dynamics, and ponderomotive oscillations in the tunneling regime, as well as single-photon excitations in thin nanojunctions. 

We classify electron transport regimes in the strong-field interaction, all of which are based on tunneling phenomena. In the photon-assisted tunneling regime, a bound electron absorbs one or more photons but remains below the threshold, tunneling through the junction barrier. Due to the presence of the image potential, the number of absorbed photons is constrained by the laser field strength and the gap width. In the multiphoton regime ($\gamma\gg1$), a bound electron absorbs multiple photons beyond the threshold and tunnels through the junction barrier. In the non-adiabatic regime ($\gamma\sim1$), the barrier is both suppressed and driven by the laser field. The electron can tunnel through the barrier while absorbing photons, causing an energy shift under the tunneling barrier. The freed electron outside the tunneling exit will be accelerated by the laser field until it transmits into the opposite metal. In the tunneling (or adiabatic) regime ( $\gamma\ll1$), the laser field strongly suppresses the junction barrier, resulting in rapid tunneling without any energy change.

We also introduce a new parameter, $\zeta$, which characterizes the interactions in MIM junctions in addition to the Keldysh parameter $\gamma$. For conditions where $\zeta\geq1$, the MIM junction will not exhibit laser-dominated photoemission, even under an intense laser field with $\gamma\ll1$. Beyond the original definition of $\zeta$ using times, this parameter has several more physical interpretations: it can represent the energy capacity ratio, which characterizes the maximum energy transferred to the electron by the laser field, and it can also be interpreted as the tunnel exit ratio under adiabatic conditions. The interaction and photoemission processes in the MIM junctions must be described jointly by the Keldysh parameter and $\zeta$.

For future experiments and attosecond observations, we examine the responses of current rectification to CEP and pulse duration. A laser pulse comprising less than three cycles is required, which remains a challenging task, dependent on the advancement of technology. In the current theory, we treat the MIM junction within a one-dimensional model, and the unidirectional trajectories as well as recollisions are not included. A more comprehensive model, incorporating a three-dimensional framework, optical near-field enhancement and many-electron interactions, remains elusive. Furthermore, the effect of bias voltage warrants further exploration. However, even the present form of our theory with its shortcomings is expected to be useful in guiding experiments and interpreting results, ultimately opening the door to attosecond-angstr\"om microscopic observations and petahertz electron transport.

\begin{acknowledgments}
We thank Andrey Borisov and Zhaopin Chen for insightful discussions. This project has received funding from the European Union's Horizon 2020 research and innovation program under grant agreement No 853393-ERC-ATTIDA and from the Israel Science Foundation under grant agreement No 1504/20. The authors also acknowledge the Helen Diller Quantum Center at the Technion for partial financial support.
\end{acknowledgments}

\appendix

\section{Averaged net current}\label{A}
The continuity relation ensures that the time-integrated current flux is conserved everywhere. Therefore, to calculate the total current, we can only focus on the current flux towards the sample side. Integrating Eq.~\ref{eq2.2.1} over the time and the sample space ($x \in (d,+\infty)$), we derive
\begin{equation}
\begin{split}\label{eqA1}
\int_{-\infty}^{+\infty}J_{s}\;dt\,=-\left\langle \Psi(t) \vert\chi_{s}\vert  \Psi (t) \right\rangle\bigg\vert^{t=\infty}_{t=-\infty}.
\end{split}
\end{equation}
Here, we use $-\infty$ and $+\infty$ to represent the initial and final times. The operator $\chi_{s}$ denotes the space integration performed in the sample region $(d,+\infty)$. The notation $[...]\big\vert^{t=\infty}_{t=-\infty}$ at the end represents the subtraction of the term evaluated at the initial time from the term evaluated at the final time. Before interacting with light, the electron is in its initial state $\left\vert  \Psi (-\infty) \right\rangle=\left\vert  \Psi_{s} \right\rangle$. The interaction bifurcates $\left\vert  \Psi_{s} \right\rangle$ into a tunneling wavefunction $\left\vert  \Psi_{sT} \right\rangle$ in the tip and a reflected wavefunction $\left\vert  \Psi_{sR} \right\rangle$ in the sample. For a wave function initially located at the tip $e^{i\phi}\left\vert \Psi_{t} \right\rangle$, it can contribute to $e^{i\phi}\left\vert \Psi_{tT} \right\rangle$ and $e^{i\phi}\left\vert \Psi_{tR} \right\rangle$, respectively. A random global phase $\phi$ is introduced, since these wavefunctions are originally from different sources (tip and sample). Hence, the wavefunction inside the sample at the final time is $\left\vert  \Psi (+\infty) \right\rangle=\left\vert  \Psi_{sR} \right\rangle+e^{i\phi}\left\vert \Psi_{tT} \right\rangle$. Considering $N$ laser pulses, the average current per pulse is 
\begin{eqnarray}\label{eqA2}
{\cal J}&=&\frac{1}{N}\sum_{n=1}^{N}\int_{-\infty}^{+\infty}J_{s}\;dt\,\nonumber\\
&=&\frac{1}{N}\sum_{n=1}^{N}\{\langle e^{i\phi(n)}\Psi_{tT}+\Psi_{sR}\vert e^{i\phi(n)}\Psi_{tT}+\Psi_{sR}\rangle\ \nonumber\\ && \ \ \ \ \ \ \ \ \ \ -\langle \Psi_{s} \vert  \Psi_{s} \rangle\}\nonumber\\
&=&\frac{1}{N}\sum_{n=1}^{N}\{\langle e^{i\phi(n)}\Psi_{tT}+\Psi_{sR}\vert e^{i\phi(n)}\Psi_{tT}+\Psi_{sR}\rangle\nonumber\\
&&\ \ \ \ \ \ \ \ \ \ -\langle \Psi_{sT}+\Psi_{sR} \vert U(+\infty,-\infty) \times \nonumber\\
&&\ \ \ \ \ \ \ \ \ \ \ \ \  U^{\dagger}(+\infty,-\infty)\vert\Psi_{sT}+\Psi_{sR}\rangle\}\nonumber\\
&=&\frac{1}{N}\sum_{n=1}^{N}\{\langle \Psi_{tT} \vert\Psi_{tT}\rangle-\langle \Psi_{sT} \vert\Psi_{sT}\rangle\nonumber\\
&&\ \ \ \ \ \ \ \ \ \ +2\Re(e^{-i\phi(n)}\langle \Psi_{tT} \vert\Psi_{sR}\rangle)\}\nonumber\\
&=&\langle \Psi_{tT} \vert\Psi_{tT}\rangle-\langle \Psi_{sT} \vert\Psi_{sT}\rangle
\end{eqnarray}
where we have used the relation $U(+\infty,-\infty)\vert\Psi_{s}\rangle=\vert\Psi_{sT}\rangle+\vert\Psi_{sR}\rangle$. The interference term is averaged out.

\section{Tip-sample transformation}\label{B}

In this appendix, we demonstrate that $M_{\mathrm{sE}}$ can be solved in the same way as $M_{\mathrm{tE}}$. The tip boundary is located at the origin ($x=0$), while the sample boundary is located at $x=d$. Hence if we flip the whole system and translate the sample's boundary to the origin, we can relabel the tip and sample. In this transform, the electric field for the sample should have an opposite sign. Assuming that this transformation is $T$, the evolution of the wavefunction of the sample $U\vert\Psi_{s}\rangle$ now changes to $T U T^{-1} T\vert\Psi_{s}\rangle$, shortly written as $U'\vert\Psi'_{s}\rangle$. 
The new tunneling amplitude is 
\begin{eqnarray}\label{eqB1}
{\langle \Psi'_{sT} \vert\Psi'_{sT}\rangle}&=&\langle \Psi'_{s} \vert U'^{-1}\chi_{s}U'\vert\Psi'_{s}\rangle\nonumber\\
&=&\langle \Psi_{s} \vert T^{-1} T U^{-1} T^{-1} \chi_{s} T U T^{-1} T\vert\Psi_{s}\rangle\nonumber\\
&=&\langle \Psi_{s} \vert U^{-1} T^{-1} T\chi_{t}T^{-1} T U \vert\Psi_{s}\rangle\nonumber\\
&=&(\langle\Psi_{sT}\vert+\langle\Psi_{sR}\vert) \chi_{t} (\vert\Psi_{sT}\rangle+\vert\Psi_{sR}\rangle)\nonumber\\
&=&\langle \Psi_{sT} \vert\Psi_{sT}\rangle
\end{eqnarray}
where $\chi_{t}$ means we only select the position domain inside the tip. Hence, the transformation does not change the tunneling amplitude.

\section{Tunneling amplitude derivations}\label{C}

To obtain the time-dependent Bardeen's tunneling matrix element from the Dyson series Eq.~\ref{eq2.3.4} we use the following two Schr\"odinger equations,
\begin{equation}\label{eqC1}
\langle\psi(t)\vert(-i\hbar\frac{\partial}{\partial t}-H_{0})[\chi_\mathrm{t}+\chi_\mathrm{s}]=
\langle\psi(t)\vert V(t)[\chi_\mathrm{t}+\chi_\mathrm{s}].
\end{equation}
and
\begin{equation}\label{eqC2}
[\chi_\mathrm{t}+\chi_\mathrm{s}](i\hbar\frac{\partial}{\partial t}-H_{0})\vert\Psi_\mathrm{Is}(t)\rangle=[\chi_\mathrm{t}+\chi_\mathrm{s}]V_\mathrm{Is}(t)\vert\Psi_\mathrm{Is}(t)\rangle.
\end{equation}
and insert them into Eq.~\ref{eq2.3.4}. 
The inner product in space representation is executed with partial integration in the following calculation:
%\small
\begin{widetext}\label{eqC3}
\begin{eqnarray}
&&\left(-\frac{i}{\hbar}\right)\int_{-\infty}^{t}\langle\psi(t_1)\vert[\chi_\mathrm{t}+\chi_\mathrm{s}][V(t_{1})-V_\mathrm{Is}(t_{1})]\vert\Psi_\mathrm{Is}(t_{1})\rangle\;dt_{1}\,\nonumber\\
&=&\int_{-\infty}^{t}\int_{-\infty}^{\infty}[\chi_\mathrm{t}+\chi_\mathrm{s}]\left[\Psi_\mathrm{Is}(x,t_1)\left(-\frac{\partial}{\partial t_1}-\frac{i\hbar}{2m}\frac{\partial^2}{\partial x^2}\right)\psi^*(x,t_1)-\psi^*(x,t_1)\left(\frac{\partial}{\partial t_1}-\frac{i\hbar}{2m}\frac{\partial^2}{\partial x^2}\right)\Psi_\mathrm{Is}(x,t_1)\right]\;dx\,\;dt_{1}\,\nonumber\\
&=&-\langle\psi(t)\vert[\chi_\mathrm{t}+\chi_\mathrm{s}]\vert\Psi_\mathrm{Is}(t)\rangle-\frac{i\hbar}{2m}\int_{-\infty}^{t}\int_{-\infty}^{\infty}[\chi_\mathrm{t}+\chi_\mathrm{s}] \frac{\partial}{\partial x}\left[\Psi_\mathrm{Is}(x,t_{1})\frac{\partial}{\partial x}\psi^{*}(x,t_{1})-\psi^{*}(x,t_{1})\frac{\partial}{\partial x}\Psi_\mathrm{Is}(x,t_{1})\right]\;dx\,\;dt_{1}\,\nonumber\\
&=&-\langle\psi_{E}(t)\vert\chi_\mathrm{s}[\chi_\mathrm{t}+\chi_\mathrm{s}]\vert\Psi_\mathrm{Is}(t)\rangle+\frac{i\hbar}{2m}\int_{-\infty}^{t} \left[\Psi_\mathrm{Is}(x,t_{1})\frac{\partial}{\partial x}\psi^{*}(x,t_{1})-\psi^{*}(x,t_{1})\frac{\partial}{\partial x}\Psi_\mathrm{Is}(x,t_{1})\right]\bigg\vert^{x=d}_{x=0}\;dt_{1}\,\\
&=&-\langle\psi_{E}(t)\vert\chi_\mathrm{s}\vert\Psi_\mathrm{Is}(t)\rangle+\frac{i\hbar}{2m}\int_{-\infty}^{t} \left[\Psi_\mathrm{Is}(x,t_{1})\frac{\partial}{\partial x}\psi^{*}(x,t_{1})-\psi^{*}(x,t_{1})\frac{\partial}{\partial x}\Psi_\mathrm{Is}(x,t_{1})\right]\bigg\vert^{x=d}_{x=0}\;dt_{1}\,.\nonumber
\end{eqnarray}
\end{widetext}

%\normalsize
This first term can exactly eliminate the time-boundary state. Substituting the above result into Eq.~\ref{eq2.3.4}, we finally obtain the tunneling amplitude in Eq.~\ref{eq2.3.5}.

\section{Volkov propagator}\label{D}

The Volkov wavefunction describes the motion of a free charged particle in a homogeneous electric field. In the length gauge, the wavefunction with a certain canonical momentum $p$ is 
\begin{equation}\label{eqD1}
\langle x\vert p \rangle=\frac{1}{\sqrt{2\pi\hbar}}e^{\frac{i}{\hbar}(p-eA(t))x}e^{-\frac{i}{\hbar}\int_{-\infty}^{t}\frac{(p-eA(\tau))^2}{2m}d\tau\,}\;.
\end{equation}
The Volkov propagator is obtained by integrating over the momentum:
\begin{eqnarray}\label{eqD2}
&&\langle x_{2}\vert U(t_{2},t_{1})\vert x_{1} \rangle=\int{\langle x_{2}\vert p\rangle\langle p\vert x_{1}\rangle}\;dp\,\nonumber\\&=&\sqrt{\frac{m}{2\pi i \hbar(t_{2}-t_{1})}}e^{\frac{i}{\hbar}(\tilde p-eA(t))x_{2}-\frac{i}{\hbar}(\tilde p-eA(t))x_{1}}\nonumber\\ &&\times e^{-\frac{i}{\hbar}\int_{-t_{1}}^{t_{2}}\frac{(\tilde p-eA(\tau))^2}{2m}\;d\tau\,}.
\end{eqnarray}
where $\tilde p=\frac{\int_{t_{1}}^{t_{2}}eA(\tau)\;d\tau\,+m(x_{2}-x_{1})}{(t_{2}-t_{1})}$. The Gaussian integral has been used.

\section{Transition amplitude}\label{E}

For simplicity, we define $\vert\tilde p-e A(t_{1})\rangle=e^{\frac{i}{\hbar}(\tilde p-e A(t_{1})) x}$, the initial wavefunction is $e^{-\frac{i}{\hbar}E_{0}t}\vert\Psi_{0}\rangle$, and the dipole operator is $-e{\cal E}(t_{1})x$. The transition amplitude in Eq.~\ref{eq4.1.1} is 

\begin{eqnarray}\label{eqE1}
\eta(t_{1},t_{2})&=&[p_{\mathrm{s}}+\tilde p-eA(t_2)](e^{-\frac{i}{\hbar}p_{\mathrm{s}}d}-R e^{\frac{i}{\hbar}p_{\mathrm{s}}d})\nonumber\\ &&\times\{-e {\cal E}(t_{1})\langle \tilde p-e A(t_{1})\vert x \vert\Psi_{0}\rangle\}.
\end{eqnarray}

\section{Photoemission formulae}\label{F}
From Eq.~\ref{eq4.1.2}, we define $S(t_{2},t_{1})$ as follows:
\begin{eqnarray}\label{eqF1}
S(t_{2},t_{1})&=&E t_{2}+\frac{{\tilde p}^2}{2m}(t_{2}-t_{1})-\int_{t_{1}}^{t{2}}\frac{e^2A^2(\tau)}{2m} \;d\tau\,\nonumber\\
&&-\int_{t_{1}}^{t{2}}V_{\mathrm{imag}}[x(\tau)]\;d\tau\,+\vert E_{0}\vert t_{1}.
\end{eqnarray}
For $\zeta<1$, we have $\Im{t_{\mathrm{1s}}}=\frac{\ln(\gamma+\sqrt{1+\gamma^2})}{\omega}$, $\Im{t_{\mathrm{2s}}}=0$, and $\Im{\tilde p_{\mathrm{(s)}}}\approx0$. Substitute these solutions into $S$, we obtain:
\begin{eqnarray}\label{eqF2}
\Im{\big[E t_{\mathrm{2s}}}\big]=\Im{\bigg[\frac{{\tilde p_{\mathrm{(s)}}}^2}{2m}(t_{\mathrm{2s}}-t_{\mathrm{1s}})}\bigg]=0,
\end{eqnarray}

\begin{eqnarray}\label{eqF3}
\Im{\bigg[\int_{t_{\mathrm{1s}}}^{t{\mathrm{2s}}}\frac{e^2A^2(\tau)}{2m} \;d\tau\,\bigg]}=\frac{\vert E_{0}\vert_{\mathrm{eff}}}{2\omega\gamma^2}\times\nonumber\\
(\gamma\sqrt{1+\gamma^2}-\omega\Im{t_{\mathrm{1s}}}),
\end{eqnarray}
and
\begin{eqnarray}\label{eqF4}
\Im{\bigg[\int_{t_{1}}^{t{2}}V_{\mathrm{imag}}[x(\tau)]\;d\tau\,\bigg]}=\overline{\vert V_{\mathrm{imag}}\vert}\Im{t_{\mathrm{1s}}}.
\end{eqnarray}

Assembling all terms, the imaginary part of the action is given by
\begin{equation}\label{eqF5}
\Im{S}=\frac{\vert E_{0}\vert_{\mathrm{eff}}}{\omega}\bigg[\ln(\gamma+\sqrt{1+\gamma^2})\frac{1+2\gamma^2}{2\gamma^2}-\frac{\sqrt{1+\gamma^2}}{2\gamma}\bigg].
\end{equation}
Eq.~\ref{eq4.2.6}, Eq.~\ref{eq4.2.7}, and Eq.~\ref{eq4.2.8} can be derived using the corresponding approximations of $\gamma$.

For $\zeta>1$, the static field is dominant. We have the duration 
\begin{equation}\label{eqF6}
\tau=\frac{-i2md}{\sqrt{2m|E|_{\mathrm{eff}}}+\sqrt{2m|E_{0}|_{\mathrm{eff}}}}.
\end{equation}
Substituting it into the three saddle-point equations Eq.~\ref{eq4.1.4}-\ref{eq4.1.6}, we obtain the relation
\begin{equation}\label{eqF7}
\cos[\omega(t_{\mathrm{1s}}+\frac{\tau}{2})]=\frac{-|E|_{\mathrm{eff}}+|E_{0}|_{\mathrm{eff}}}{|e|Fd}.
\end{equation}
It has an extremum at $|E|_{\mathrm{eff}}=|E_{0}|_{\mathrm{eff}}-|e|Fd$. The corresponding time at the extremum is 
\begin{equation}\label{eqF8}
\Im{t_{\mathrm{1s}}}=-\Im{t_{\mathrm{2s}}}=\frac{md}{\sqrt{2m|E|_{\mathrm{eff}}}+\sqrt{2m|E_{0}|_{\mathrm{eff}}}}.
\end{equation}
Substituting them into the action $S$ and expanding $S$ to third order, we obtain
\begin{eqnarray}\label{eqF9}
S(t_{\mathrm{2s}},t_{\mathrm{1s}})&=&d\sqrt{2m|E_{0}|_{\mathrm{eff}}}\bigg[1-\frac{|e|Fd}{4|E_{0}|_{\mathrm{eff}}}\nonumber\\&&-\frac{(|e|Fd)^2}{32|E_{0}|_{\mathrm{eff}}^2}-\frac{(|e|Fd)^3}{128|E_{0}|_{\mathrm{eff}}^3}\bigg]\nonumber\\&\approx& d\sqrt{2m(|E_{0}|_{\mathrm{eff}}-|e|Fd/2)}.
\end{eqnarray}
Eq.~\ref{eq4.2.5} comes from this result.

\section{Comparison between different propagators}\label{G}

\begin{figure*}[htb!]
\centering
\includegraphics[clip, trim=0cm 5.5cm 0.5cm 4cm, width=1\linewidth]{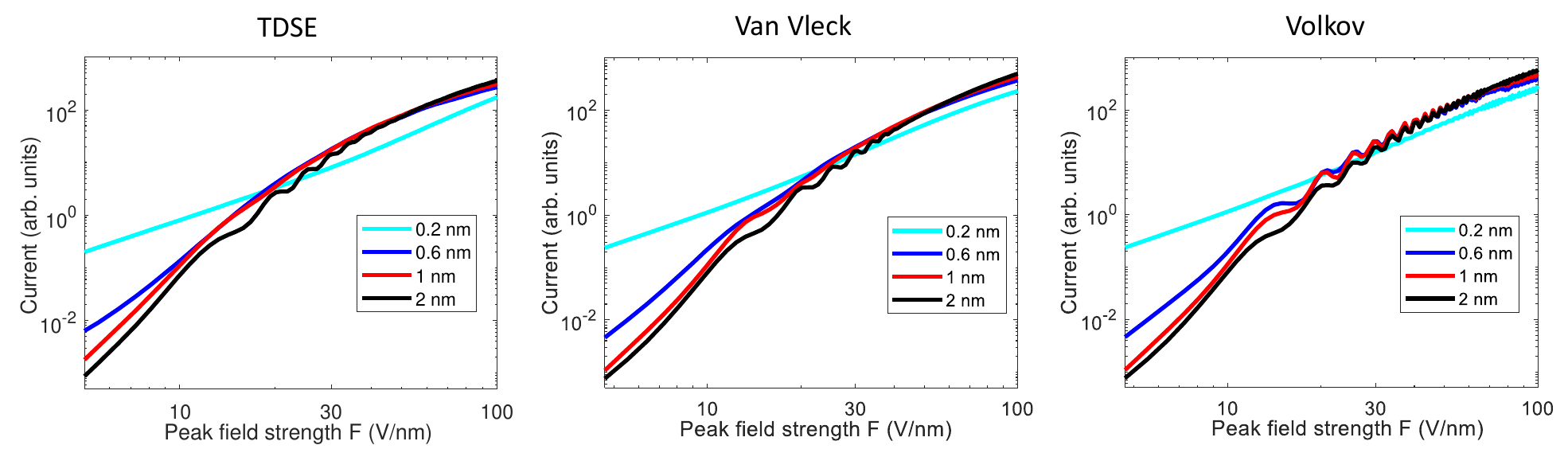}
\caption{Comparisons between Van Vleck and Volkov calculations are presented, with all parameters matching those in Fig.~\ref{Fig5}(a). Compared to the numerical results of the TDSE, the Van Vleck method provides highly accurate results. In contrast, the Volkov calculations overestimate the ponderomotive energy and channel closing due to unconstrained trajectories, leading to rapid oscillations beyond 17 V/nm. }
\label{FigS1}
\end{figure*}

Figure~\ref{FigS1} presents a comparison of calculations based on Van Vleck and Volkov propagators with TDSE results.

%
% FIGURE S1
%

%\bibliography{scibib}  % Produces the bibliography via BibTeX.


\end{document}